\documentclass[12pt,fleqn]{article}
\usepackage{bezier,amsmath,amsfonts}
\usepackage[dvips]{graphics}
\usepackage{hyperref}
\usepackage{float}

\usepackage{amsfonts,amssymb,cite}
\usepackage{graphicx}

\def\onecol{\onecolumn \mathindent 2em}

\def\noi{\noindent}

\newcommand{\Title}[1]{\noi {{\Large\bf #1}}\\[1ex]}

\def\Aunames#1{\noi{\bf #1}}
\def\au#1{${}^{#1}$}
\def\Addresses#1{\medskip\noi \protect
	\begin{description}\itemsep -3pt {\it #1} \end{description}}
\def\adr#1#2{\item[${}^{#1}$]{\it #2}}

\newcommand{\Abstract}[1]{\vskip 2mm \begin{center}
        \parbox{16.4cm}{\small\noi #1} \end{center}\medskip}

\def\email#1#2{\footnotetext[#1]{e-mail: #2}\addtocounter{footnote}{1}}

\def\nq{\hspace*{-1em}}
\def\nqq{\hspace*{-2em}}
\def\nhq{\hspace*{-0.5em}}

\def\qq{\qquad}
\def\cm{\hspace*{1cm}}

\def\wide{\mbox{$\dst\vphantom{\int}$}}

\usepackage{color}

\def\Acknow#1{\subsection*{Acknowledgments} #1}

\def\Jl#1#2{#1 {\bf #2},\ }

\def\ApJ#1 {\Jl{Astroph. J.}{#1}}
\def\CQG#1 {\Jl{Class. Quantum Grav.}{#1}}
\def\DAN#1 {\Jl{Dokl. AN SSSR}{#1}}
\def\GC#1 {\Jl{Grav. Cosmol.}{#1}}
\def\GRG#1 {\Jl{Gen. Rel. Grav.}{#1}}
\def\IJMPD#1 {\Jl{Int. J. Mod. Phys. D}{#1}}
\def\JETF#1 {\Jl{Zh. Eksp. Teor. Fiz.}{#1}}
\def\JETP#1 {\Jl{Sov. Phys. JETP}{#1}}
\def\JHEP#1 {\Jl{JHEP}{#1}}
\def\JMP#1 {\Jl{J. Math. Phys.}{#1}}
\def\NPB#1 {\Jl{Nucl. Phys. B}{#1}}
\def\NP#1 {\Jl{Nucl. Phys.}{#1}}
\def\PLA#1 {\Jl{Phys. Lett. A}{#1}}
\def\PLB#1 {\Jl{Phys. Lett. B}{#1}}
\def\PRD#1 {\Jl{Phys. Rev. D}{#1}}
\def\PRL#1 {\Jl{Phys. Rev. Lett.}{#1}}

\def\al{&\nhq}
\def\lal{&&\nqq {}}
\def\eq{Eq.\,}
\def\eqs{Eqs.\,}
\def\beq{\begin{equation}}
\def\eeq{\end{equation}}
\def\besub{\begin{subequations}}
\def\esub{\end{subequations}}
\def\bear{\begin{eqnarray}}
\def\bearr{\begin{eqnarray} \lal}
\def\ear{\end{eqnarray}}
\def\earn{\nonumber \end{eqnarray}}
\def\nn{\nonumber\\ {}}
\def\nnv{\nonumber\\[5pt] {}}
\def\nnn{\nonumber\\ \lal }
\def\nnnv{\nonumber\\[5pt] \lal }
\def\yy{\\[5pt] {}}
\def\yyy{\\[5pt] \lal }
\def\eql{\al =\al}

\def\dst{\displaystyle}
\def\tst{\textstyle}

\def\fract#1#2{{\tst\frac{#1}{#2}}}

\def\half{{\fract{1}{2}}}

\def\e{{\,\rm e}}
\def\d{\partial}

\def\im{\mathop{\rm Im}\nolimits}

\def\sign{\mathop{\rm sign}\nolimits}
\def\diag{\mathop{\rm diag}\nolimits}

\def\const{{\rm const}}
\def\eps{\varepsilon}

\def\then{\ \Rightarrow\ }

\newcommand{\vars}[1]{\left\{\begin{array}{ll}#1\end{array}\right.}

\newcommand{\Picture}[3]{
	\begin{figure} \centering \unitlength=1mm
	\begin{picture}(130,#1)
		\put(0,0){\line(0,1){#1}}       
		\put(0,0){\line(1,0){130}}              
		\put(130,0){\line(0,1){#1}}             
		\put(0,#1){\line(1,0){130}}
	\put(0,0){#2}                       \end{picture}
        \caption{\protect\small #3}  \medskip \hrule \end{figure} }

\def\eqn#1{\eq\eqref{#1}}
\def\rf{\eqref}

\def\M{{\mathbb M}}
\def\ME{\mbox{$\M_{\rm E}$}}
\def\MJ{\mbox{$\M_{\rm J}$}}

\def\R{{\mathbb R}}
\def\Z{{\mathbb Z}}

\def\og{{\overline g}}

\def\oR{{\overline R}}
\def\oom{{\overline \omega}}
\def\oq{{\overline q}{}}

\def\df{\delta\phi}
\def\da{\delta\alpha}
\def\db{\delta\beta}
\def\dg{\delta\gamma}
\def\dpsi{\delta\psi}

\def\kappa{\varkappa}
\def\Geff{G_{\rm eff}}
\def\Veff{V_{\rm eff}}
\def\tall{\mbox{$\tst\vphantom{\int^0}$}}

\def\GR{general relativity}
\def\pb{perturbation}
\def\pbs{perturbations}
\def\sph{spherically symmetric}
\def\ssph{static, spherically symmetric}
\def\asflat{asymptotically flat}
\def\wh{wormhole}
\def\whs{wormholes}
\def\bh{black hole}
\def\bhs{black holes}

\def\emag{electromagnetic}
\def\grav{gravitational}
\def\Scw{Schwarz\-schild}
\def\Schr{Schr\"o\-din\-ger}
\def\RN{Reiss\-ner-Nord\-str\"om}

\def\umx{u_{\max}}

\def\mn{_{\mu\nu}}
\def\MN{^{\mu\nu}}
\def\mN{_\mu^\nu}

\usepackage{color}

\allowdisplaybreaks
\begin{document}
\onecol

\Title{On the stability of electrovacuum space-times\yy in scalar-tensor gravity}

\Aunames{Kirill A. Bronnikov,\au{a,b,c,1} Sergei V. Bolokhov,\au{b,2} Milena V. Skvortsova,\au{b,3}\\
			Rustam Ibadov,\au{d,4} and Feruza Y. Shaymanova\au{d,f,5} }
	
\Addresses{\small
\adr a	{Center of Gravitation and Fundamental Metrology, VNIIMS, 
		Ozyornaya ul. 46, Moscow 119361, Russia}
\adr b	{Institute of Gravitation and Cosmology, RUDN University, 
		ul. Miklukho-Maklaya 6, Moscow 117198, Russia}
\adr c 	{National Research Nuclear University ``MEPhI'', 
		Kashirskoe sh. 31, Moscow 115409, Russia}
\adr d {Department of Theoretical Physics and Computer Science, Samarkand State University, 
		Samarkand 140104, Uzbekistan}
\adr e {Karshi State  University, Kochabog street, Karshi city 180103, Uzbekistan}
		}

\Abstract
   {We study the behavior of static, spherically symmetric solutions to the field equations of  
   scalar-tensor theories (STT) of gravity belonging to the Bergmann-Wagoner-Nordtvedt class,
   in the presence of an electric and/or magnetic charge. 
   This class of theories includes the Brans-Dicke, Barker and Schwinger STT as well as 
   nonminimally coupled scalar fields with an arbitrary parameter $\xi$.  The study is restricted 
   to canonical (nonphantom) versions of the theories and scalar fields without a self-interaction 
   potential.  Only radial (monopole) perturbations are considered as the most likely ones to cause an 
   instability. The static background solutions contain naked singularities, but we formulate
   the boundary conditions in such a way that would preserve their meaning if a singularity is 
   smoothed, for example, due to quantum gravity effects. These boundary conditions look 
   more physical than those used by other authors.  Since the solutions of all STT under 
   study are related by conformal transformations, the stability problem for all of them reduces 
   to the same wave equation, but the boundary conditions for perturbations (and sometimes 
   the boundaries themselves) are different in different STT, which affects the stability results. 
   The stability or instability conclusions are obtained for different branches of solutions in 
   the theories under consideration and are presented in a table form.
    }

\email 1 {kb20@yandex.ru} 
\email 2 {boloh@rambler.ru} 
\email 3 {milenas577@mail.ru} 
\email 4 {ibrustam@mail.ru}
\email 5 {yusupovnafiruz89@gmail.com}

\section{Introduction}

  Stability studies are known to be an important part of theoretical physics and a necessary 
  stage in the analysis of static or stationary solutions aimed at a description of any real or
  thinkable object, including stars, black holes, wormholes and other configurations described 
  by various theories of gravity.

  Of particular interest are such studies for solutions to \grav\ field equations involving scalar fields. 
  Unlike vector (\emag) and tensor \grav\ \pbs, those of a scalar field contain a monopole degree 
  of freedom that can most likely lead to an instability of an isolated (in particular, \asflat) 
  field configuration. The reason is that, at least in the case of spherical symmetry, the effective 
  potential $\Veff$ in the master equation for \pbs\ contains a term of the form $\ell(\ell+1)/r^2$ 
  or like that, creating what may be called a centrifugal barrier. On the other hand, the master
  equation with proper boundary conditions leads to a boundary-value problem whose 
  eigenvalues are squared perturbation frequencies $\Omega^2$, and if $\Omega^2 < 0$, we 
  have to conclude an instability of the original static or stationary solution because there are \pbs\   
  that grow exponentially with time. And, in full similarity to the problems in quantum mechanics
  (where the eigenvalues represent energy levels), positive contributions like $\ell(\ell+1)/r^2$ to 
  $\Veff$ always lead to an increase in the eigenvalues. Therefore, if a system under consideration is 
  unstable, this instability will be most probably indicated at the smallest multipolarity $\ell$,
  that is, for monopole perturbations if they exist.    

  Most of the stability studies of configurations involving scalar fields concern such objects as \bhs, 
  \whs\ and boson stars (see, e.g., \cite{kon-zh11, trap17, stab18, br-book, brito23} and referenced therein).
  Less attention was paid to scalar solutions with naked singularities, despite the fact that many
  if not the majority of solutions to gravitational field equations with scalar fields contain such
  singularities. Such solutions indeed seem unplausible due to the popular cosmic censorship 
  conjecture, and moreover, there is an ambiguity in formulating reasonable boundary conditions 
  for \pbs\ at such singularities. In our opinion, any singularities, be they naked or not, cannot be
  plausible since one can hardly believe in the existence of infinite curvatures or densities 
  in Nature (evident exceptions are weak singularities invoked at idealized descriptions of 
  topological defects, strings, branes, thin shells, etc.). Such infinities are most likely suppressed
  or smoothed in more advanced theories or by quantum gravity effects. A regularization is
  then expected to only touch upon tiny regions replacing the singularities, while outside 
  them remain areas well described by classical dynamics. Can we expect that our stability
  studies, carried out for singular solutions, will preserve their relevance in regularized solutions? 
  We suppose that the answer is yes if the boundary conditions for \pbs\ are chosen so as to 
  survive a future regularization. That is the approach we follow in the present paper.  
  
  Probably the first stability study carried out in this spirit was \cite{kb-hod}, it concerned 
  Fisher's \cite{fisher} \ssph\ solution with a massless minimally coupled scalar field in general 
  relativity (GR) (much later re-discovered by Janis, Newman and Winicour \cite{JNW} 
  as well as by Wyman \cite{wyman81} and sometimes called the JNW solution) and its 
  counterpart with an electric charge obtained by Penney \cite{penney}. 
  It was concluded in \cite{kb-hod} that these solutions are unstable under radial perturbations. 
  This study may be easily extended to scalar-tensor theories (STT) of gravity, belonging to the 
  Bergmann-Wagoner-Nordtvedt class \cite{STT1, STT2, STT3}, since their solutions are 
  connected with those of GR with a minimally coupled scalar by a well-known conformal 
  transformation \cite{STT2}. These theories belong to the most widely discussed alternatives 
  to GR, able to solve many important problems of astrophysics and cosmology.
  Nevertheless, to our knowledge, an extension of the stability study to their \sph\ solutions 
  was not considered for many years. For scalar-vacuum solutions it was discussed in our 
  recent paper \cite{we23}.
	
  Contrary to \cite{kb-hod}, Sadhu and Suneeta \cite{sadhu13} came to the conclusion on
  a ``naked singularity stable under scalar field perturbations'' in Fisher's (JNW) space-time,
  and later on, on this basis, quasinormal frequencies of Fisher's singularity were  calculated
  \cite{chow20}. However, the \pbs\ discussed in \cite{sadhu13} represent a test scalar field, 
  while we discuss \pbs\ of the scalar that supports the system itself and take into account the 
  accompanying metric \pbs, therefore, we are dealing with different stability problems, the one 
  considered in \cite{sadhu13} being manifestly incomplete. In addition, the boundary condition
  for a test scalar field at the naked singularity were formulated in \cite{sadhu13} on the basis 
  of such mathematical requirements as a self-adjoint nature of the Sturm-Liouville problem
  and square integrability of the solution and were different from ours, but the latter are,
  in our view, more physical in nature. A more complete study by Clayton et al. \cite{clayton98}
  indicates that ``space-times with vanishing potential are unstable,'' thus confirming the 
  result of \cite{kb-hod}, but there is ``a branch of the space-times with non-vanishing
  scalar field mass which is perturbatively stable.'' Further results on the \pbs\ and stability of 
  space-times in GR supported by self-interacting scalar fields with a sufficiently wide range
  of scalar field potentials have been presented by Stashko et al. in \cite{sta23, sta24a}, and for 
  those in quadratic $f(R)$ gravity in \cite{sta24b}.

  In the present paper, we continue the study begun in \cite{we23} and consider 
  electrovacuum solutions of the same (Bergmann-Wagoner-Nordtvedt) class of STT, conformally 
  related to Penney's space-time, thus restricting ourselves to STT with massless scalar fields.
  We can remark that such solutions play in these theories the same role as the \RN\ solution in 
  GR and, in our opinion, definitely deserve a stability study. A question of interest is whether 
  an electric charge can stabilize such configurations that decay being electrically neutral.
  
  We will consider monopole perturbations of \ssph\ space-times \MJ\ in the framework of STT,
  comprising their Jordan conformal frame, which are conformally related to Penney's space-time 
  \ME\ that comprises their Einstein conformal frame. The conformal factors that connect them 
  depend on the choice of STT. In this paper, we restrict our study to theories with a canonical 
  behavior of the scalar field, bearing in mind that theories with a phantom behavior of the
  scalars are conformal to other branches of the extended Penney solution, and their properties 
  are quite different from those of their canonical analogues \cite{br73, stab18}). We will also 
  leave aside the emerging cases of so-called conformal continuations \cite{br73, kb-CC2}, in 
  which the whole manifold \ME\ is mapped to only a part of \MJ, and it becomes necessary to 
  extend \MJ\ to new regions with, in general, negative values of the effective gravitational 
  constant $\Geff$ \cite{kb-CC2, br-star07, skvo10}. 
  
  The structure of the paper is as follows. In Section 2 we briefly discuss the setup of the 
  general STT and some its special cases. Section 3 is devoted to \ssph\ scalar-electrovacuum 
  solutions of GR and their STT counterparts. In Section 4 we consider the equations for
  \sph\ \pbs, which have a common form for all examples of the static solutions under
  consideration. In Section 5 we consider the boundary conditions for perturbations that 
  depend on the choice of STT as well as the parameters of particular solutions and formulate
  the emerging boundary-value problems. In Section 7 we numerically solve these problems, 
  obtaining conclusions on the stability or instability of the solutions under study. Finally, 
  Section 7 is a conclusion containing a table that summarizes the main results of our study.

\section{Scalar-tensor theories}

  We will deal with the general Bergmann-Wagoner-Nordtvedt STT of gravity with
  the action  \cite{STT1, STT2, STT3}
\bearr   \label{S_J}
             S_{\rm STT} = \frac 1{16\pi} \int \sqrt{-g} d^4 x
		             \Big[f(\phi) R + 2 h(\phi)\phi^{,\alpha} \phi_{,\alpha} 
  			           - 2 U(\phi) + L_m\Big],
\ear    
  where $R$ is the scalar curvature, $g = \det(g\mn)$, $f(\phi), h(\phi)$, and $U(\phi)$ 
  are arbitrary functions ($f(\phi) > 0$ describing a nonminimal coupling between the 
  scalar field $\phi$ and the curvature), and $L_m$ is the Lagrangian of nongravitational 
  matter. This formulation of STT is known as the Jordan (conformal) frame which is 
  specified in pseudo-Riemannian space-time \MJ\ with the metric $g\mn$. Next, by the 
  well-known conformal mapping 
\beq              \label{map}
		g\mn = \og\mn/f(\phi),
\eeq  
  the theory is converted to the so-called Einstein conformal frame, specified in space-time 
  \ME\ with the metric $\og\mn$, where the action takes the form inherent to \GR\ with a 
  minimally coupled scalar field $\psi$,
\bearr   \label{S_E}
             S_{\rm STT} = \frac 1{16\pi} \int \sqrt{-\og} d^4 x
		             \Big[\oR + 2 \eps \og\MN \psi_{,\mu} \psi_{,\nu} 
  			           - 2 U(\phi)/f^2(\phi) + L_m/f^2(\phi)\Big],
\ear      
  where quantities obtained from or with $\og\mn$ are marked with a bar, while the fields 
  $\phi$ and $\psi$ are related by 
\beq  			\label{phi-psi}
		\frac {d\phi}{d\psi} = \frac{\sqrt 2 f(\phi)}{\sqrt{|D|}}, \qq
		D = fh + \frac 32 \Big(\frac{df}{d\phi}\Big)^2, \qq \eps = \sign D.
\eeq  
  In the case $\eps =1$ the field $\psi$ is of canonical nature, while if $\eps =-1$, the scalar
  $\phi$ has a negative kinetic energy and is called phantom.
  
  Evidently, if we know a solution to the field equations in $\ME$, it is easy to obtain its counterpart 
  in $\MJ$ using the transformation \rf{map} and \rf{phi-psi}, so that the line element in \MJ\ is
\beq             \label{ds_J}
		  ds_J^2 = g\mn dx^\mu dx^\nu = \frac 1{f(\phi(\psi))} ds_E^2. 
\eeq   
  
  In this paper, we will discuss the stability properties of electrovacuum solutions of GR and the 
  following examples of STT \rf{S_J}: 
\begin{enumerate}
\item  
	The Brans-Dicke theory \cite{BD-STT}:
\besub	           \label{BD}
\bearr           \label{BDa}
		f(\phi) = \phi, \qq h(\phi) = \frac{\omega}{\phi}, \qq \omega = \const \ne -3/2,
\\ \lal     \nq      \label{BDb}
		\then\ \psi = \psi_0  + \frac {\oom}{\sqrt 2} \log |\phi|, \quad\
		\phi = \e^{\sqrt 2 (\psi-\psi_0)/\oom}, \quad\
		\psi_0=\const, \quad\   \oom = \sqrt{|\omega + 3/2|}.
\ear
\esub
\item
	Barker's theory \cite{barker}, in which the effective gravitational constant is really a constant:
\besub	           \label{Bark}
\bearr            \label{Bark-a}
		f(\phi) = \phi, \qq h(\phi) = \frac{4-3\phi}{2\phi(\phi-1)},
\yyy            \label{Bark-b}
	    \then\ \psi = \psi_0 + \arctan \sqrt{\phi -1},\qq	\phi = \frac 1 {\cos^2 (\psi-\psi_0)},
	    \quad\ \psi_0=\const, 
\ear
\esub	
   where we have assumed $\phi >1$, corresponding to a canonical nature of $\psi$.   
\item
	Schwinger's theory \cite{schwg, bruk94}: 
\besub	           \label{Swg}
\bearr            \label{Swg-a}			
		f(\phi) = \phi, \qq h(\phi) = \frac{K-3\phi}{2\phi^2}, \qq K = \const > 0,
\yyy 			  \label{Swg-b}		
		\then\   \psi = \psi_0 + \sqrt{K/\phi},	\qq	   \phi = \frac K {(\psi-\psi_0)^2}.
    	     \quad\ \psi_0=\const, 
\ear
\esub	
\item
	The STT often considered as GR with a nonminimally (conformally or nonconformally)
	coupled scalar field:
\beq                \label{nonmin}
		f(\phi) = 1 - \xi \phi^2, \qq  h(\phi) = 1, \qq \xi = \const.	
\eeq	
     This theory splits into four cases with different expressions for $\psi(\phi)$, to be discussed
     later on: (a) $\xi = 1/6$ (conformal coupling),  (b) $0 < \xi < 1/6$, (c) $\xi > 1/6$, 
     and (d) $\xi < 0$.  
\end{enumerate}  
  In what follows, the indices $E$ and $J$ will mark quantities belonging to \ME\ and \MJ,
  respectively.

\section{Static scalar-electrovacuum solutions of GR and STT}
\subsection{Solutions in \ME: Derivation}

  We will consider static background \sph\ space-times, beginning with those of GR, 
  or, which is the same, in the Einstein frame \ME\ of the STT \rf{S_J}.
  We thus deal with the action \rf{S_E} where $U(\phi) =0$ and the Maxwell \emag\ 
  field Lagrangian $L_m = -F\mn F\MN$,  
\bear                                                  \label{S-GR}
     S _E =\frac 1{16\pi} \int {\sqrt{-\og}} \Big(\oR 
	         + 2\eps \og^{\alpha\beta}\psi_{,\alpha}\psi_{,\beta}- F\mn F\MN\Big),
\ear
  and, as usual, $F\mn = \d_\mu A_\nu - \d_\nu A_\mu$. Note that the gravitational 
  constant $G$ is here absorbed in the field definitions.

  The general \sph\ metric in \ME\ may be written in the form
\beq                                                         \label{ds1}
    ds_E^2 = \og\mn dx^{\mu}dx^{\nu}
           = \e^{2\gamma} dt^2 - \e^{2\alpha}du^2 - \e^{2\beta}d\Omega^2,
\eeq
  where $\gamma$, $\alpha$ and $\beta$ are functions of the radial coordinate $u$ 
  and time $t$, and $d\Omega^2= d\theta^2 + \sin^2\theta\,d\varphi^2$. 
  We will also use the notation for the spherical radius $r(u,t) \equiv \e^\beta$.
  A center (if any) corresponds, by definition, to $r \to 0$.
  
  Let us present the nonzero components of the Ricci tensor, preserving only linear
  terms with respect to time derivatives (for using in the subsequent stability study):  
\bear                                                  \label{Rmn}
	R_0^0 \eql  \e^{-2\gamma}(2\ddot{\beta} + \ddot{\alpha})
			 -  \e^{-2\alpha} [\gamma'' + \gamma'(2\beta' + \gamma' -\alpha')];
\nnv
	R_1^1 \eql \e^{-2\gamma} \ddot{\alpha}
       - \e^{-2\alpha} [2\beta'' + \gamma'' + 2\beta'^2 + \gamma'^2
              -\alpha' (2\beta' + \gamma')];
\nnv
       R_2^2 \eql \e^{-2\beta} + \e^{-2\gamma} \ddot{\beta} 
       - \e^{-2\alpha} [\beta'' + \beta' (2\beta' + \gamma' -\alpha')] = R_3^3;
\nnv
       R_{01} \eql 2 [\dot{\beta}{}' + \dot{\beta}\beta'
      			          - \dot{\alpha}\beta' - \dot{\beta} \gamma'],
\ear
  where dots and primes stand for $\d/\d t$ and $\d/\d u$, respectively.
  
  In static configurations to be considered in this section, we are dealing with only
  $u$-dependent quantities $\og\mn$ and $\psi$; moreover, the \emag\ potential $A_\mu$
  may only have nonzero components $A_0$, describing a Coulomb electric field,  and 
  $A_3 = \oq \cos\theta$ responsible for a monopole magnetic field, where $\oq$ is the 
  magnetic charge. Let us, for certainty, restrict ourselves to an electric field. Consideration of 
  a magnetic charge or both charges simultaneously will not change the results of this study: 
  we would only have to replace $q$ in all formulas with $\sqrt{q^2 + \oq^2}$. 

  Then the electromagnetic field equations lead to
\beq                                                         \label{F10}
	\e^{2\alpha} F^{10} = q,\qq  q = \text{electric charge}. 
\eeq
  We thus have $F\mn F\MN = -2q^2 \e^{-4\beta}$.
  The scalar field equation $\Box \psi =0$ yields
\beq           \label{psi'}
        \e^{-\alpha + 2\beta +\gamma}\psi' = C, \qq C = \text{scalar charge}.
\eeq
  For convenience, without loss of generality, we will everywhere assume $C > 0$.
  
  The Einstein equations $G\mN \equiv \oR\mN - \half \delta\mN \oR = - 8\pi T\mN$ are 
  solved in elementary functions in a general form \cite{br73}, since the stress-energy tensors 
  (SETs) of the scalar and \emag\ fields have the simple structure
\bear	 						\label{SET-chi}
	8\pi T\mN [\phi] \eql \eps \e^{-2\alpha} \phi'^2
			\diag (1,\ -1,\ 1,\ 1),
\\                                                      \label{SET-F}
	8\pi T\mN [F] \eql q^2\e^{-4\beta}\diag (1,\ 1,\ -1,\ -1).
\ear

  Thus, in the case $\phi \equiv \const$ we have the \RN\ (RN) solution with the
  metric (in curvature coordinates, $u=r$, with $m$ being the \Scw\ mass.)
\beq           \label{RN}
      ds_E^2 = A(r) dt^2 - dr^2/A(r) -r^2 d\Omega^2,          \qq
      A(r) = 1 - \frac{2m}{r} + \frac{q^2}{r^2},
\eeq

  In the case $\phi \ne \const$ we have a more general solution which was first 
  found by Penney \cite{penney} for $\eps =+1$ and extended to $\eps=-1$ in \cite{br73}.
  It may be called the extended Penney solution. Let us reproduce it, following \cite{br73},
  in terms of the harmonic coordinate $u$, chosen according to the condition
\beq
        \alpha(u) = 2\beta(u) +\gamma(u).                              \label{harm}
\eeq
  In these coordinates, since for our system $T^1_1 + T^2_2 =0$, the corresponding 
  combination of the Einstein equations reads $(\beta+\gamma)'' = \e^{2\beta+ 2\gamma}$
  (a Liouville equation) and is easily integrated giving
\beq                           \label{beta}
    	\e^{-\beta-\gamma} = s(k,u) \equiv
     		 \vars{ k^{-1}\sinh\,ku,  & k > 0; \\
     				       u, & k = 0;  \\
     			   k^{-1}\sin ku, & k < 0, }  
\eeq
  where $k=\const \in \R$, and one more integration constant has been eliminated by 
  shifting the origin of $u$. Consequently, with no loss of generality, one can assert that 
  the harmonic $u$ coordinate is defined for $u > 0$, and $u = 0$ corresponds to spatial 
  infinity. The metric can now be written as
\beq                                                            \label{ds2}
	ds_E^2 = \e^{2\gamma} dt^2 - \frac{\e^{-2\gamma}}{s^2(k,u)}
			  \biggl [\frac{du^2}{s^2(k,u)} + d\Omega^2\biggr].
\eeq
  By (\ref{beta}), at small $u$ ($u\to 0$), assuming $\gamma(0) =0$, the conventional 
  flat-space spherical radial coordinate $r=\e^{\beta}$ is connected with $u$ by $u = 1/r$.

  Note that in the metric (\ref{ds2}) it is easy to pass over to the convenient ``quasiglobal'' 
  coordinate $x$ specified by the condition $\alpha + \gamma =0$ in (\ref{ds1}), but this 
  transition is different for different $k$. Namely:
\begin{itemize}
\item
	At $k > 0$, we put $\e^{-2ku} = 1 - 2k/x$, and the metric takes the
	form
  \beq                                                        \label{ds+}
	ds_E^2 = \e^{2\gamma} dt^2 - \e^{-2\gamma} \biggl[ dx^2
		     + x^2 \biggl(1 - \frac{2k}{x}\biggr) d\Omega^2\biggr].
  \eeq
\item
	At $k = 0$, we put $u = 1/x$ to obtain
  \beq                                                        \label{ds0}
       ds_E^2 = \e^{2\gamma} dt^2 - \e^{-2\gamma}
				          (dx^2 + x^2 d\Omega^2).
  \eeq
\item
	At $k < 0$, we put $|k| \cot (|k|u) = x$ and arrive at
  \beq                                                        \label{ds-}
	ds_E^2 = \e^{2\gamma} dt^2 - \e^{-2\gamma}
				\Big[ dx^2 + (x^2 + k^2) d\Omega^2\Big].
  \eeq
\end{itemize}

  The remaining unknown $\gamma$ is found from the Einstein equation 
  $\oR{}^t_t = ...$ which, in the coordinates (\ref{harm}), has again a Liouville form,
  $\gamma'' = q^2 \e^{2\gamma}$, from which $\gamma(u)$ is easily found:
  specifically, we obtain $\e^{- \gamma(u)} = |q| s(h, u+u_1)$ where $h$ and $u_1$ 
  are integration constants, and the function $s(h, u+u_1)$ is defined similarly to \rf{beta},
  or more explicitly,
\beq                                         \label{gamma}
    	\frac 1{|q|} \e^{-\gamma} =  s(h,u + u_1) \equiv
     		 \vars{h^{-1}\sinh\,h(u+u_1),  & h > 0; \\
     				       u + u_1, & h = 0;  \\
     			   h^{-1}\sin h(u+u_1), & h < 0.}  
\eeq  
   Also, in the coordinates \rf{harm}, \eqn{psi'} gives simply $\psi = Cu$ 
  (assuming $\psi(0) =0$ without loss of generality). The whole solution takes the form
\bear
     ds_E^2 \eql                                                  \label{ds_E}
         \frac{dt^2}{q^2\,s^2(h,u+u_1)} -
     				 \frac{q^2\,s^2(h,u+u_1)}{s^2(k,u)}
          \biggr[\frac{du^2}{s^2(k,u)} + d\Omega^2\biggl],
\yy
     F_{\mn} \eql                                               \label{F_mn}
	     (\delta_{\mu 0}\delta_{\nu 1}
		-\delta_{\nu 0}\delta_{\mu 1})\,q \e^{\alpha+\gamma-2\beta}
	     = (\delta_{\mu 0}\delta_{\nu 1}
	    -\delta_{\nu 0}\delta_{\mu 1})\,\frac{1}{q\,s^2(h,u+u_1)};
\\
     \psi \eql  Cu.                                              \label{psi}
\ear
  Lastly, the Einstein equation $G^u_u =\ldots$, being a first integral of the system, 
  leads to the following relation constraining the integration constants $C,\ k,\ h$:
\beq
              k^2\sign k = \eps C^2 + h^2\sign h .               \label{int}
\eeq

  The range of $u$ is $0 < u < \umx$, where $u=0$ corresponds to spatial infinity, 
  while $\umx$ may be finite or infinite depending on the constants $k,$ $h,$ $u_1$; 
  thus, $\umx = \infty$ if $k\geq 0$ and $h\geq 0$, and also $u_1 > 0$. 
  The metric is \asflat\ at $u=0$ (see \eqs \rf{ds+}--\rf{ds-} to make it evident), and 
  if we accordingly require $\og_{tt}(0) =1$, this constrains the integration constant $u_1$ by
\beq
			s^2(h,\ u_1) = 1/q^2                                   \label{u_1}
\eeq
  (preserving some discrete arbitrariness of $u_1$ in the case $h<0$).  We thus have three 
  essential integration constants: either $k$ or $h$, and the charges $q$ and $C$. An 
  expression for the  mass $m$ of the configuration is obtained by comparing the asymptotic 
  behavior of \rf{ds_E} at small $u \approx 1/r$ with the Schwarzschild metric: 
\beq
	       m = \pm \sqrt{q^2 + h^2 \sign h}.      \label{m}
\eeq
  For $h\geq 0$, one has $\sign m = \sign u_1$, and for $h< 0$, $\sign m = \sign \sin (|h|u_1)$. 
  It seems natural to require $m \geq 0$. However, since in STT solutions with metrics conformal to 
  \rf{ds_E}, the corresponding expressions for the mass $m_J$ will be different from \rf{m}, 
  in what follows we will admit any nonzero values of $u_1$.

  The RN solution is restored when $C=0 \then h=k$. Evidently, in this case, $h =k > 0$ 
  corresponds to a non-extremal RN \bh, $h= k = 0$ to an extremal \bh, and $h=k < 0$ to
  the RN solution with a naked singularity.
  
  If there is no \emag\ field, $q=0$, then the Einstein equation $\oR{}^t_t = ...$ in 
  the coordinates (\ref{harm}) reads simply $\gamma'' =0$, which leads to the metric  
  \rf{ds2} with $\e^{2\gamma} = \e^{-2h u}$ and the relation between the constants
  $ k^2\sign k = \eps C^2 + h^2$ instead of \rf{int}; the \Scw\ mass $m$ is equal to $h$. 
  It is Fisher's well-known solution \cite{fisher} in the case $\eps =+1$ and its phantom 
  counterpart \cite{ber-lei} if $\eps=-1$. These space-times have been studied in detail, 
  including their stability properties, see, e.g., 
  \cite{h-ell73, br73, kb-hod, br-book, sarb1, stab11, stab18}, 
  and the stability of their STT counterparts for $\eps=+1$ was recently discussed in 
  \cite{we23}, so here we will focus on configurations with $q \ne 0$.  

\subsection{Solutions in \ME: Classification}

   Solutions of the family (\ref{ds_E})--(\ref{psi}) may be classified as follows:
\bear
	[1+] &&  \eps=+1,\ \ k > h >0 ;    \nn
	[2+] &&  \eps=+1,\ \ k > h =0 ;    \nn
	[3+] &&  \eps=+1,\ \ h<0 ;           \nn
	 [1-] &&  \eps=-1,\ \ h > k >0 ;      \nn
	 [2-] &&  \eps=-1,\ \ h > k =0 ;      \nn
	 [3-] &&  \eps=-1,\ \ h \geq 0,\ k <0 ; \nn
	 [4-] &&  \eps=-1,\ \ 0 > h > k .     \label{class}
\earn
  Let us briefly characterize each class according to its behavior near $u=\umx$.
\begin{description}
\item[{$[1+]$, $u_1 >0:$}]
       $\umx = \infty$ is an attracting singular center ($r=\e^{\beta}\to 0$)
       since there $\e^{\gamma} \sim \e^{-hu} \to 0$. It is a scalar-type
       singularity, similar to that in Fisher's solution and, moreover, the total 
       scalar field energy $E_s$ in the whole space-time is infinite whereas  
       the electric field energy $E_e < \infty$.
\item[{$[2+]$, $u_1 >0:$}]
       The same as [1+] but $\e^{\gamma} \sim 1/u$ as $u\to\infty$.
\item[{$[3+]:$}]
       $\umx$ is determined by the nearest zero of $\sin [|h|(u+u_1)]$,
       i.e., $\umx = \pi/|h| - u_1$. This is a repulsive, RN type
       singularity, where $\e^{\gamma}\to \infty$; the total electric field
       energy $E_e$ is infinite whereas $E_s < \infty$. 
\item[{$[1+],\ [2+]$, $u_1 <0:$}]       
       The same behavior as in class $[3+]$ but now $\umx = - u_1$.
\item[{$[1-]$, $u_1 >0$}]:
       $\umx = \infty$, where $\e^{\gamma} \sim \e^{-hu} \to 0$ and
       $r = \e^\beta \sim \e^{(h-k)u}\to \infty$. A configuration without a
       curvature singularity which was previously called a ``cold \bh''
       since it has, by construction, a zero Hawking temperature. Its
       horizon $u=\infty$ has an infinite area. The solution can, however,
       be extended beyond such a horizon only in the cases $h = nk$, $n =
       2,3,\ldots$. (This is actually a special case of cold \bhs\
       considered in Ref.\,\cite{cold-q}).
\item[{$[2-]$, $u_1 >0:$}]
       The same as [1-], but with a different functional dependence of
       $r=\e^{\beta (u)}$. There is no analytical extension of the metric
       beyond the surface $u=\infty$.
\item[{$[3-{\rm o}]$}, $u_1 >0:$]
       $\umx = \pi/|k|$ is another spatial infinity, where $\gamma$ tends to
       a finite limit while $r=\e^\beta \to \infty$. The whole configuration
       is a static, \asflat, traversable \wh. 
\item[{$[3-]$}, $u_1 < 0:$]       
       In the cases $h\geq 0$, $u_1 <0$, the qualitative nature of
       the geometry depends on which of the functions $\sinh[h(u+u_1)]$ 
       (simply $u=u_1$ if $h=0$) or $\sin |k|u$ will be first to reach zero,
       see Fig.\,1, left panel.  Three types of behavior are possible:
\begin{figure*}
\centering
\includegraphics[width=8cm]{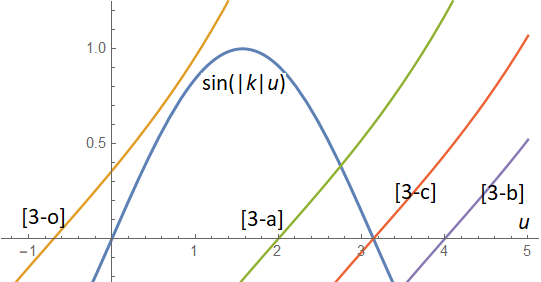}\qq
\includegraphics[width=7cm]{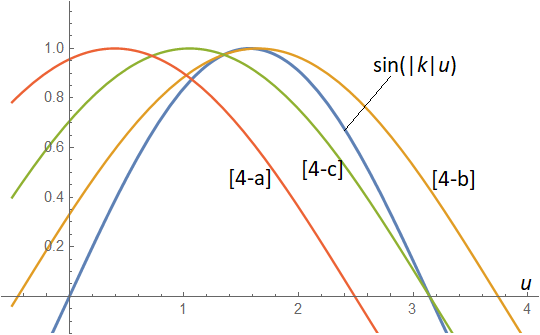}
\caption{\small
	Solutions $[3-]$ (left) and $[4-]$ (right): different positions (a,b,c) of the
	curves $\sinh[|h|(u+u_1)]$ in the left panel and $\sin[|h|(u+u_1)]$ 
	in the right panel with respect to $\sin(|k|u)$ determine the nature of geometry.
	In particular, it is a \wh\ if $\umx$ is determined by only $\sin(|k|u)$, and 
	$\umx = \pi/|k|$	is then the second flat spatial infinity (the first one is at $u=0$). }
\end{figure*}
    
\medskip\noi
     [3-a]: $\umx = -u_1 < \pi/|k|$; the solution behaves as that of class $[3+]$
     (an RN type singularity).

\medskip\noi
     [3-b]: $\umx = \pi/|k| < -u_1$: a behavior like that of class [3-o] with $u_1 > 0$
     (a \wh).

\medskip\noi
     [3-c]: $\umx = - u_1 = \pi/|k|$. As $u\to\umx$,
     $\e^{\gamma}\to\infty$, while $\e^{\beta}$ and the Kretschmann scalar
     tend to finite limits. So we obtain a singularity-free hornlike
     structure (like the ones obtained in some solutions of dilaton gravity
     \cite{horn}), with an infinitely remote (since $l=\int \e^{\alpha}du$
     diverges) ``end of the horn'', whose radius $r =\e^{\beta}$ is
     asymptotically constant and repels test particles.

\item[{$[4-]$}]:
       $\umx = \min \{ \pi/|k|,\ \pi/|h|-u_1\}$. The qualitative nature of
       the geometry is determined by the interplay of the two sines: $\sin
       (|k|u)$ and $\sin [|h|(u+u_1)]$ as shown in Fig.\,1, right panel. Depending 
       on the positions of their zeros, we obtain three cases 4-a, 4-b, 4-c 
       with properties quite similar to those of class 3-a, 3-b, 3-c with $u_1 < 0$.
\end{description}

     The diagram in Fig.\,2 presents a map of the whole family of solutions (\ref{ds_E})--(\ref{psi}) 
     for a fixed charge $q$ in the case $m >0$. The notations are
\beq
      \mu = m^2/q^2, \cm \sigma = \eps C^2/q^2.             \label{not-f2}
\eeq

 The separatrix between wormhole solutions and those with a RN naked
  singularity is the curve at which $\pi/|h|-u_1 = \pi/|k|$, or, in terms
  of $\mu$ and $\sigma$,
\beq                                                          \label{sepa}
	-\sigma = (1-\mu) \Biggl[1
		- \frac{\pi^2}{(\pi-\arcsin\sqrt{1-\mu})^2}\Biggr].
\eeq

\Picture{72}
{
\unitlength=.72mm
\special{em:linewidth 1pt}
\linethickness{1pt}
\begin{picture}(147.00,129.00)(7,37)
\put(20.00,50.00){\vector(1,0){150.00}}
\put(100.00,50.00){\vector(0,1){80.00}}
\put(20.00,80.00){\line(1,0){150.00}}
\put(100.00,80.00){\line(-5,6){40.00}}
\bezier{400}(25.00,50.00)(100.00,55.00)(100.00,80.00)
\put(125.00,46.00){\makebox(0,0)[cc]{1}}
\put(100.00,46.00){\makebox(0,0)[cc]{0}}
\put(75.00,46.00){\makebox(0,0)[cc]{-1}}
\put(50.00,46.00){\makebox(0,0)[cc]{-2}}
\put(25.00,46.00){\makebox(0,0)[cc]{-3}}
\put(102.00,81.00){\makebox(0,0)[lb]{1}}
\put(101.00,131.00){\makebox(0,0)[lc]{$\mu = m^2/q^2$}}
\put(160.00,44.00){\makebox(0,0)[cc]{$\sigma = \eps C^2/q^2$}}
\put(119.00,108.00){\makebox(0,0)[cc]{$[1+]$}}
\put(140.00,81.00){\makebox(0,0)[cb]{$[2+]$}}
\put(127.00,68.00){\makebox(0,0)[cc]{$[3+]$}}
\put(91.00,56.00){\makebox(0,0)[cc]{[4--a]}}
\put(54.00,67.00){\makebox(0,0)[cc]{[4--b]}}
\put(26.00,81.00){\makebox(0,0)[cb]{$h=0$}}
\put(55.00,131.00){\makebox(0,0)[cc]{$k=0, [2-]$}}
\put(89.00,109.00){\makebox(0,0)[cc]{$[1-]$}}
\put(54.00,95.00){\makebox(0,0)[cc]{$[3-]$}}
\end{picture}
}
{Map of extended Penney solutions with a fixed charge $q\ne 0$ in the plane 
 ($\mu, \sigma$), where $\mu = m^2/q^2$ and $\sigma =\eps C^2/q^2$.
 The \RN\ solution occupies the axis $\sigma=0$. Solutions with a RN-like naked singularity
 occupy the band $0 < \mu < 1$ to the right of the separatrix curve, which
 itself depicts solutions of class [4--c]. Class [$2-$] solutions are
 situated on the line $k=0$.}
 
\subsection{STT solutions}

  As follows from Section 2, \ssph\ solutions of an arbitrary STT in its Jordan frame \MJ\ are
  obtained from those in \ME\ (\rf{ds_E}--\rf{psi} in the general case) using the conformal 
  mapping \rf{map} and the transformation \rf{phi-psi} for the scalars. The \emag\ field 
  $F\mn$ is the same in both frames due to the conformal invariance of the \emag\
  Lagrangian $\sqrt{-g}F\mn F\MN$. Due to this conformal invariance, the Maxwell field in 
  \MJ\ possesses the same electric-magnetic duality as in \ME, and the constant $q$ in all 
  solutions may be understood not only as an electric charge but as a ``mixture'' of electric 
  and magnetic charges. 
  
  The qualitative features of the metric remain the same after the mapping \rf{map} 
  if the conformal factor $f(\phi)$ is smooth and finite in the whole range of $\phi$ 
  (or $\psi$) including its boundaries. As we will see in particular examples, it is usually 
  not the case, and the solution in \MJ\ is drastically different from its counterpart in \ME. 
  
  This difference is especially important in those cases where a singularity in \ME\ maps 
  to a regular surface in \MJ, and the latter should then be continued through this surface,
  implementing a so-called conformal continuation, a phenomenon studied for \ssph\ 
  space-times in \cite{kb-CC1, kb-CC2}. In such cases, the whole manifold \ME\ maps 
  to only a part of \MJ. It is clear that the stability problem should be then formulated
  separately from the generic situation where the map \rf{map} puts the 
  points of \ME\ and \MJ\ in one-to-one correspondence. In some STT, for example, in  
  the Barker and Schwinger theories, it happens, on the contrary, that only a part of \ME\   
  maps into the whole \MJ\ which terminates due to a singularity in the conformal factor.
  In such cases, we can obviously use the \pb\ equations formulated in \ME\ but impose 
  the boundary conditions on a surface corresponding to the edge of \MJ.

  In the present paper, we will study the stability of solutions belonging to the canonical 
  sector of STT ($\eps = +1$), hence those of branches $[1+]$, $[2+]$ and $[3+]$ in \ME,
  excluding the possible cases of conformal continuations. Evidently, such a continuation due 
  to \rf{map} is only possible if different metric coefficients in \ME\ (that is, $\e^{2\gamma}$ 
  and $\e^{2\beta} \equiv r^2$ in our metric \rf{ds1}) turn to zero or infinity in the same 
  manner, thus admitting their simultaneous ``correction'' by a properly behaved conformal 
  factor. In the solution \rf{ds_E}--\rf{psi} this happens only in branch $[1+]$ in the special 
  case $k=2h$, hence $C^2 = 3h^2 = 3(m^2 - q^2)$.  It is certainly only a necessary 
  condition since in order to cure a singularity the conformal factor must exhibit a precisely 
  opposite behavior. In what follows we will mention some such cases. A stability study for
  phantom STT solutions is even more laborous than for the canonical sector and is 
  thus far postponed for the future.
  
  The solutions in \MJ\ to be considered are quite diverse, and we will briefly discuss their 
  properties further on in Section 5 where we are going to formulate the relevant boundary
  conditions for their \pbs.  
  
\section{Perturbation equations}  
  
  Let us now subject the solutions in \ME\ to \sph\ perturbations: suppose that there is a 
  static solution with the metric \rf{ds1}, some $\psi = \psi(u)$ and a radial electric field, 
  consider a perturbed function
\[
    \psi(u,t)= \psi(u)+ \dpsi(u,t)
\]
  and similarly introduce \pbs\ of the metric functions $\da,\ \db,\ \dg$. 
  In such a case, there is only a single dynamic degree of freedom related to the scalar field
  \pbs\ because the gravitational and \emag\ perturbations cannot be \sph\ (monopole). 
  Therefore, the analysis of perturbed field equations must lead to a single wave equation 
  in terms of $\dpsi$, and its study must answer the question on stability or instability of the 
  background static configuration. It turns out that this wave equation is most easily 
  derived using the perturbation gauge $\db \equiv 0$ (this corresponds to the choice of 
  a particular reference frame in the nonstatic perturbed space-time). 
  In \cite{stab11, stab18, br-book} it has been shown that the resulting wave equation 
  is gauge-invariant and therefore describes the behavior of real perturbations of 
  our system rather than possible coordinate effects. 
  
  The derivation starts from writing the scalar field equation $\Box \psi =0$
  in the metric \rf{ds1} using an arbitrary radial coordinate $u$ but in the linear 
  approximation with respect to time derivatives:
\beq
		2\e^{2\alpha-2\gamma} \delta\ddot\psi - 2[\dpsi'' +\dpsi' (\gamma'+2\beta'-\alpha') 
          +\psi'(\dg' + 2\db' -\da')]  =0.
\eeq  
  Next, assuming $\db \equiv 0$, we can express the difference $\dg' - \da'$ in terms of 
  $\da$ and the background quantities using the Einstein equation $R^2_2 = \ldots$ 
  (see \rf{Rmn}), and lastly $\da$ is expressed in terms of $\dpsi$ using the Einstein equation
  $R_{01} = - T_{01}$ which takes an especially simple form with $\db =0$: 
\beq
			\beta'\delta{\dot\alpha} = \eps\psi' \delta\dot\psi
			\ \ \then\ \ \beta'\da = \eps \psi'\dpsi,
\eeq  
  where we have neglected the emerging arbitrary function of $u$ since we focus on nonstatic \pbs. 
  As a result, we obtain the following wave equation for $\dpsi$ in terms of an arbitrary 
  radial coordinate $u$  \cite{stab11, stab18, br-book, kb-hod}:
\bearr  			\label{wave-eq}
      \e^{2\alpha-2\gamma} \delta\ddot\psi  -\dpsi'' 
      - \dpsi' (\gamma'+ 2\beta'-\alpha') + W(u) \dpsi =0,
\yyy                     \label{W}
      W(u) \equiv  \frac{2\eps \psi'{}^2}{\beta'^2}\e^{2\alpha -2\beta}\big(q^2 \e^{-2\beta} -1\big)
		    \equiv  \frac{2\eps  \e^{2\alpha} \psi'{}^2}{r'{}^2}\Big(\frac{q^2}{r^2} -1\Big);
\ear  
  recall that we have $\eps = 1$ for a canonical scalar field $\psi$ under consideration. 
  Since our background solution is static, we can separate the variables in a usual manner by 
  putting 
\beq  
 			\dpsi = \e^{i\Omega t} X(u),\qq \Omega = \const,
\eeq  
  leading to an ordinary differential equation for $X(u)$,
\beq  			\label{eq-X}
            X'' + (\gamma'+ 2\beta'-\alpha') X' 
        					+ [\e^{2\alpha-2\gamma}\Omega^2 - W(u)] X =0,
\eeq  
  which may be called the {\bf master equation} for linear perturbations of our background system.
  The next step is to transform \eqn{eq-X} to a canonical form by substituting 
\beq 			\label{X_to_Y}
		X(u) = \e^{- \beta} Y(z),
\eeq  
  where, as before, $\beta = \log r$ is taken from the static solution, while
  $z$ is the so-called tortoise coordinate related to an arbitrary radial coordinate $u$ 
  in \rf{ds_E} by 
\beq                \label{u_to_z}
       \frac {du}{dz} = \e^{\gamma (u)-\alpha(u)}, \qq z = \int \e^{\alpha(u) -\gamma(u)} du.     
\eeq  
  At the transition \rf{u_to_z}, let us choose the integration constant so that $z=0$ 
  corresponds to $r = 0$. As a result, we obtain a \Schr-like equation for $Y(z)$ 
  \cite{kb-hod, stab18}:
\beq                                                        \label{Schr}
      \frac {d^2 Y}{dz^2} + [\Omega^2 - \Veff(z)] Y =0,
\eeq
  where the effective potential $\Veff$ is expressed in terms of an arbitrary radial coordinate 
  $u$ as
\beq  			\label{Veff-u}
  	\Veff(u) = \e^{2\gamma-2\alpha}\big[W(u)+ \beta'' + \beta'(\beta'+\gamma'-\alpha')\big].
\eeq  

  Considering our scalar-electrovacuum solutions with the metric \rf{ds_E}
  (and even with scalar-vacuum solutions studied previously \cite{kb-hod, we23}), 
  a problem is that $z(u)$ obtained from \rf{u_to_z} is in all cases quite a long expression 
  depending on the solution parameters and containing combinations of hypergeometric 
  functions. It is impossible to solve it for $u(z)$ and hence to write $\Veff$ as an explicit 
  function of $z$. Therefore, \eqn{Schr} can be used for some qualitative inferences and 
  asymptotic analysis, but it cannot be solved exactly, and so it is more reasonable to 
  obtain and analyze numerical solutions to \eqn{eq-X} written in terms of the coordinate 
  $u$. Here we will begin the discussion with the asymptotic behavior of $X(u)$ and $Y(z)$.
  
  First of all, at small $u$ (corresponding to large radii $r \approx z \approx 1./u$), we find that 
  in all cases $[1+]$, $[2+]$ and $[3+]$, $\Veff \approx 2m u^3 \approx 2m/z^3$, 
  and the approximate behavior of all solutions to \eqn{Schr} reads
\beq               \label{Y0+}
			Y(z) \approx C_1 \e^{|\Omega|z} + C_2 \e^{-|\Omega|z}
\eeq
  if $\Omega^2 < 0$ (corresponding to a possible exponential \pb\ growth, 
  $\dpsi \sim \e^{|\Omega|t}$), and
\beq               \label{Y00}
			Y(z) \approx C_3 + C_4 z
\eeq  
   if $\Omega = 0$ (corresponding to a possible linear \pb\ growth, $\dpsi \sim t$); 
   here and henceforth all $C_i$ are constants. 
   
   At the other end of the range of $u$, in all solutions under consideration we have 
   $z\to 0$. It is easy to find the asymptotic form of $z(u)$ at small $z$ and the 
   corresponding behavior of the potential $\Veff$. With the metric \rf{ds_E}, the 
   transition \rf{u_to_z} and the expression \rf{Veff-u} for $\Veff$ give for the three 
   branches of the solution:
\bearr  \nhq      \label{u-inf}
	[1+], [2+], u_1 >0:\qq 
		k > h \geq 0, \ u\to \infty, \qq
		 z \approx \dfrac{\e^{2\beta}}{2(k-h)} \to 0, \qq   \Veff \approx -\dfrac 1{4 z^2}.
\yyy   \nhq       \label{u-fin}		 
	[1+], [2+], u_1 < 0, {\rm and}\ [3+]: \ \  
		u \to u_s < \infty,\quad\ \
		z \approx \dfrac 13 \e^{2\beta} (u_s\! - \!u)\to 0, \quad \Veff \approx - \dfrac{2}{9 z^2}.  
\ear
\begin{figure*}
\centering
\includegraphics[width=5.7cm]{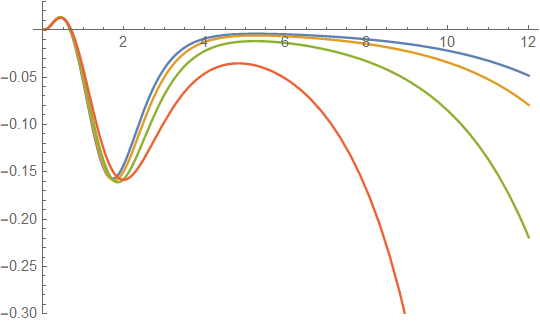}\ 
\includegraphics[width=5.7cm]{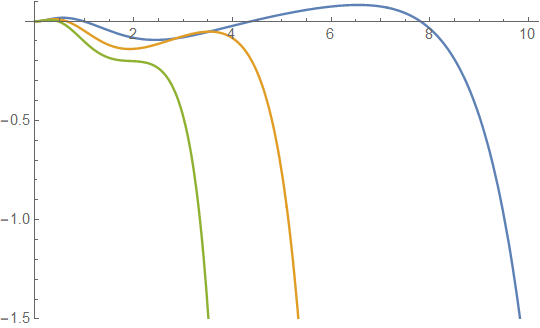}\
\includegraphics[width=5.7cm]{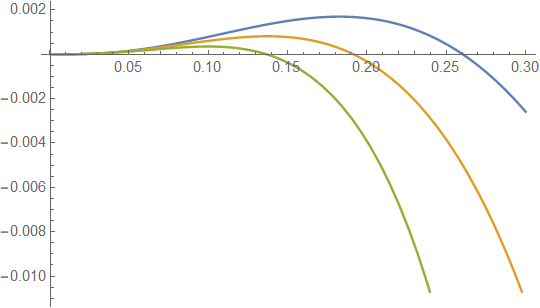}\\
\cm (a) \hspace{5cm} (b) \hspace{5cm} (c)
\caption{\small
	Examples of the behavior of $\Veff(u)$ for branches $[1+]$, $[2+]$ and $[3+]$ of 
	the solution: \\ (a) branch $[1+]$, $m=1,\ C=0.45,\ q = 0.2,\ 0.4,\ 0.6,\ 0.8$;
	(b) branch $[2+]$, $m=q=1,\ C=0.3,\ 0.5,\ 0.7$;	
	(c) branch  $[1+]$, $m=1,\ C=1,\ q = 1.5,\ 2,\ 2.5$
	(ordered upside down in all cases).
	}
\end{figure*}
  
  It is then straightforward to find the general solution to \eqn{Schr} near $z=0$ with any 
  $\Omega$ (since any finite $\Omega^2$ is negligible against the infinite negative potential): 
  thus, for the cases \rf{u-inf} we obtain
\beq 				\label{Y1,Y2}
		Y(z) \approx \sqrt z (C_5 + C_6 \log z), \qq z\to 0, \qq C_5, C_6 = \const,
\eeq  
  while for the cases \rf{u-fin}
\beq                  \label{Y3}
           Y(z) \approx C_7 z^{1/3} + C_8 z^{2/3}, \qq z\to 0, \qq C_7, C_8 = \const.
\eeq  

  The boundary conditions for \pbs\ and the subsequent stability inferences are 
  different for solutions of different STT, and we will discuss them further on.
  
  Figure 3 shows some plots of the function $\Veff(u)$ for all three branches of the solution;
  at other values of the parameters, $\Veff$ behaves qualitatively in a similar way. 
 
  It is quite possible to study the stability problem using the master equation  \rf{eq-X}
  with any suitable coordinate $u$ (and it will actually be done below in terms of the 
  harmonic coordinate that eliminates the term with $X'$ in \eqn{eq-X}). However, the canonical 
  form \rf{Schr} of the \pb\ equation written in terms of the tortoise coordinate $z$, has the
  advantage that it corresponds to the classical wave operator $\d_{tt} - \d_{zz}$, and the form of 
  the corresponding potential $\Veff$ allows us to make some qualitative conclusions or predictions 
  which are impossible with $W(u)$. Thus, since $\Veff \to -\infty$ as $z\to 0$ ($u\to\infty$),
  we can expect negative eigenvalues $\Omega^2$ by analogy with quantum mechanics.
  On the other hand, since $\Veff > 0$ at small $u$ (hence large $z$), we can expect the absence
  of eigenvalues $\Omega^2 <0$ if the relevant region is restricted to large $z$. Even though 
  our boundary conditions are different from those in quantum mechanics, these qualitative 
  predictions, achievable by watching Fig.\,3, are actually justified by the calculations, while in 
  some cases the instability conclusions are obtained directly using the asymptotic expressions
  \rf{Y1,Y2} and \rf{Y3} treated in terms of $z$.
  
\section{Boundary conditions and the stability problem}  
  
  Equation \rf{eq-X} or \rf{Schr} can be used to analyze the stability of background static 
  solutions under monopole (\sph) perturbations. Thus, if we find a nontrivial solution to \rf{eq-X} 
  or \rf{Schr} with $\im \Omega < 0$, satisfying physically meaningful boundary conditions at 
  the ends of the range of $u$ or $z$ where the solution is specified 
  (in particular, assuming absence of ingoing waves), then it can be concluded that the 
  static background system is unstable since the field perturbation $\dpsi$ can exponentially 
  grow with time. If such a solution exists with $\Omega =0$, there is an instability due to a
  possible linear growth of \pbs\ with time. If such solutions are absent, we can conclude that
  the system is stable under \sph\ perturbations in the linear approximation.
  
  From the standpoint of differential equations, the transformation \rf{map} is nothing else than 
  a substitution in the field equations, therefore, \eq \rf{Schr} (or \rf{eq-X}), obtained in the 
  \ME\ formulation of the theory, actually describes \pbs\ in an arbitrary STT; however, as long
  as we interpret the Jordan frame \MJ\ as the one physically distinguished, we must formulate 
  the boundary conditions in \MJ, and they will be different within different STT. Let us discuss 
  such boundary conditions and their consequences for the theories enumerated in Section 2.
  
  To begin with, in all solutions under consideration, the background fields are well-behaved
  at flat spatial infinity, where $\psi = Cu \sim 1/r \approx 1/z$, hence it is reasonable to set 
  a common boundary condition for all these STT, including, in particular, Penney's solution of
  GR (or in \ME), that is, 
\beq               \label{bc-infty}
		  |X|/u \sim |X| z < \infty, \qq  |Y| < \infty\ \ \ {\rm as}\ \ \ z\to\infty,    
\eeq
  since $Y(z) = r X(z) \approx z X(z)$ at large $z$. Moreover, according to \rf{Y0+}, this 
  condition actually means $C_1=0$ and $Y\to 0$ as $z \to \infty$ for $\Omega^2 < 0$, and 
  only for $\Omega =0$ it admits a finite limit of $Y$ while $C_4 =0$.
  
  At the other end, be it regular or singular, it is natural to formulate the boundary conditions 
  in terms of the scalar field $\phi$ involved in the action \rf{S_J}. There is, however, an 
  ambiguity due to reparametrization freedom inherent to STT: in \rf{S_J} one may arbitrarily 
  substitute $\phi = \phi(\Phi)$ without any change in the physical content of the theory (under 
  evident requirements to smoothness of $\phi(\Phi)$). It is therefore reasonable to formulate 
  the boundary conditions unambiguously for the function $f(\phi)$ characterizing the 
  nonminimal coupling: let us require $|\delta f/f| < \infty$. Thus, by analogy with 
  \cite{kb-hod,stab11,stab12,we23}, we impose a minimal condition compatible with the 
  perturbation scheme, admitting \pbs\ diverging not stronger than does the background field.
  
  Then, in particular, if $f(\phi)$ remains finite, we require that the \pb\ $\delta f$ should 
  be also finite, and if not, then $\delta f$ should not grow faster than $f$. Moreover, in the 
  cases where $f \to 0$, it is also reasonable to require $\delta f \to 0$. Indeed, recalling 
  that the effective gravitational constant $G_{\rm eff}$ in STT is proportional to $1/f$, 
  it appears reasonable to forbid perturbations growing faster than $G_{\rm eff}$ where
  the latter blows up. 
  
  In examples 1--3 of STT enumerated in Section 2, we have $f(\phi) = \phi$, therefore, 
  the appropriate boundary condition will simply read $|\delta \phi/\phi| < \infty$. In the 
  fourth example with $f = 1-\xi \phi^2$ we use the general rule $|\delta f/f| < \infty$. 

\subsection{Solutions and boundary conditions in particular theories} 
  
\paragraph{Penney's solution of \GR.} Let us begin with the solution \rf{ds_E}--\rf{psi}
  of GR that belongs to the class of theories \rf{S_J} with simply $f(\phi) = 1$, $\phi = \psi$ 
  and $h(\phi) =1$. At spatial infinity, as already said, we require $|Y| < \infty$. 
  
  On the other end, in branches $[1+]$ and $[2+]$  with $u_1 > 0$, we have $u \to \infty$ and 
  $\psi = Cu \to \infty$; it is a central singularity where $r = \e^\beta \to 0$ and $z \to 0$.
  Then, since at small $z$ we have $z \sim r^2$ and $\dpsi \sim X = Y/r$, to satisfy the 
  requirement $|\dpsi/\psi| < \infty$, we can admit 
\beq        \label{bc-gr1} 
		\dpsi \sim u \sim \log z \ \ \then \ \ Y(z) \sim \sqrt z \log z.
\eeq    
  Comparing \rf{bc-gr1} with \rf{Y1,Y2}, we find that all solutions to \eqn{Schr} with 
  any finite $\Omega^2$, including solutions with $\Omega^2 < 0$, satisfy our 
  boundary condition at $z=0$, and in particular, this concerns solutions that properly
  behave as $z\to \infty$ ($u \to 0$). Thus there are physically admissible perturbations 
  $\dpsi$ that grow as $\e^{|\Omega|t}$ with any value of $|\Omega|$, indicating a 
  catastrophic instability of Penney's solution belonging to these two branches. 
  
  In branch $[3+]$ as well as $[1+]$ and $[2+]$  with $u_1 < 0$, 
  we have at the other end $u \to u_s < \infty$ and a finite $\psi \to Cu_s$,
  hence we must require there  $|X| \sim |\dpsi| < \infty$. It is again a central singularity, 
  where now $r \sim u_s -u$, and $z \sim r^2(u_s-u)$, hence $r \sim z^{1/3}$, and 
\beq   			\label{bc-gr3}
         Y(z) = r X(z) \sim  z^{1/3} X(z) \ \ \then \ \ |Y(z)|/z^{1/3} < \infty.
\eeq  
  Comparing \rf{bc-gr3} with \rf{Y3}, we see again that all solutions to \eqn{Schr} with 
  any $\Omega$ satisfy our boundary condition, telling us about a catastrophic instability.
  
  Thus all three branches of Penney's solution are unstable under linear radial \pbs, 
  confirming the results obtained in \cite{kb-hod}.
      
\paragraph{The Brans-Dicke theory, \eqn{BD}.}
  The static solution has the form 
\beq                \label{sol-BD}
         f(\phi) = \phi = \e^{\sqrt 2 (\psi - \psi_0)/\oom}, \qq 
		ds_J^2 =  \e^{-\sqrt 2 (\psi - \psi_0)/\oom}ds_E^2,\qq
		\psi_0 = \const,
\eeq
  where $ds_E^2$ is given by \eqn{ds_E}. A more detailed description of these geometries 
  may be found in \cite{cold-q} in almost the same notations and in \cite{BD-q} in other notations. 
  
  In branches $[1+]$ and $[2+]$ with $u_1 > 0$, the solution is defined in the range 
  $u \in (0, \infty)$.
  
  In branch $[1+]$ ($k > h >0$) the metric coefficients in \MJ\ behave as follows at large $u$: 
\beq
		g^J_{tt} \sim \e^{2(\sigma - h)u}, \qq
		r^2_J = -g^{(J)}_{\theta\theta} \sim \e^{2(\sigma + h - k)u},\qq
		\sigma := \frac {C}{\sqrt 2\,\oom} = \frac {C}{\sqrt{2\omega+3}}.
\eeq   
  Both $g^J_{tt}$ and $r_J^2$ can have zero, finite or infinite limits, depending on the 
  interplay of the parameters $h, k$ and $\sigma$, the latter also depending on the 
  Brans-Dicke coupling constant $\omega$. In all cases, at $u \to \infty$ there is a 
  curvature singularity, with just one exception: under the conditions $k=2h$, $C=h\sqrt 3$,
  $\sigma =h$, and $\omega =0$, the surface $u=\infty$ is a regular sphere, and the 
  solution should be continued beyond it, for example, using the new variable 
  $y = \e^{-2hu}$ that may be extended to negative values. 
  This case is here excluded from consideration.
  
  At all other values of the parameters, due to \rf{sol-BD}, the condition for \pbs\  
  $|\df/\phi| < \infty$ as $u\to\infty$ leads to the requirement $|\dpsi| < \infty$,
  hence $|X| < \infty$ and consequently $|Y|/\sqrt z < \infty$ (recall that we consider 
  $\dpsi$ using quantities defined in \ME). this condition is more stringent than 
  $|\dpsi/\psi| < \infty$ obtained for Penney's solution. As a result, in the expression 
  \rf{Y1,Y2} we should require $C_6 =0$, and we arrive at a complete boundary-value 
  problem for \eqn{Schr} with the boundary conditions $|Y|< \infty$ as $z\to \infty$ and 
  $Y\sim \sqrt{z}$ as $z \to 0$. 
  
  In branch $[2+]$ ($k > h =0,\ u_1 > 0$), at large $u$ the metric in \MJ\ is characterized by 
\beq  
	  g^J_{tt} \sim \frac{\e^{2\sigma u}}{u^2}, \qq
		r^2_J = -g^{(J)}_{\theta\theta} \sim u^2 \e^{2(\sigma - k)u},\qq
		\sigma = \frac {C}{\sqrt{2\omega+3}}, \qq h = |C|.
\eeq    
  Depending on the parameter values, these quantities may tend to zero or infinity, but
  in any case $u \to \infty$ is a curvature singularity. For \pbs, as in the previous case, 
  we obtain the requirement $|\dpsi| < \infty$ and the same boundary-value problem, 
  though with another expression for the potential $\Veff$.
  
  In branch $[3+]$ ($h < 0$), as well as in $[1+]$ and $[2+]$  with $u_1 < 0$,
  the range of $u$ is $0 < u < u_s$ (either $u_s =- u_1$ or $\pi/|h| - u_1$ without 
  loss of generality), and the conformal factor $1/\phi$ is finite in this whole range, hence, 
  considering the \pbs, we are dealing with the same situation as in Penney's solution 
  and obtain the same result: a catastrophic instability under \sph\ \pbs.
  
\paragraph{Barker's theory \rf{Bark}.} We now have
\beq                   \label{sol-bark}  
		f(\phi) = \phi =  [\cos (\psi-\psi_0)]^{-2},\qq
		  ds_J^2 = \cos^2 (\psi-\psi_0)\ ds_E^2, \qq \psi_0 = \const,
\eeq  
  with $ds_E^2$ given, as before, in \eqn{ds_E}, and $\psi = Cu$.
  Assuming $\psi_0 \ne \pi/2 + n\pi, \ n \in \Z$, the solution is \asflat\ at $u=0$. 
  
  In branches $[1+]$ and $[2+]$ with $u_1 > 0$, the solution is defined in the range 
  $u \in (0, u_s)$, where $u_s$ is the closest zero of $\cos (\psi-\psi_0)$. At $u = u_s$ 
  we observe an attracting ($g^J_{tt} \to 0$) singular center ($r_J =0$) 
  
  If we require for the perturbation $|\df/\phi| < \infty$, then, as is easily verified,
  for the corresponding $\dpsi$ near the singularity $u = u_s$ we must require 
  $\dpsi = O(u - u_s)$. The radius $r$ (defined in \ME) is finite at $u= u_s$, and also
  shifts of $u$ and $z$ are of the same order since at such an intermediate value
  of $u$ the tortoise coordinate $z$ smoothly depends on $u$ according to \rf{u_to_z},
  Therefore the conditions are $|X|/(u_s - u) < \infty  \ \Leftrightarrow \ |Y|/(z - z_s) < \infty$, 
  where $z_s = z(u_s)$. The stability problem thus reduces to a boundary-value problem 
  for \eqn{Schr} on the range $(z_s, \infty)$ (corresponding to ($0, u_s$)) with the 
  conditions $|Y|< \infty$ as $z\to \infty$ and $|Y|/(z - z_s) < \infty$ as $z \to z_s$. 
  
  In branch $[3+]$ or $[1+]$ and $[2+]$  with $u_1 < 0$,
  we have precisely the same situation as just described if $u_s < |u_1|$ (in other words,
  $\cos(\psi - \psi_0)$ comes to zero faster than $s(h (u+u_1))$) and arrive at the 
  same boundary-value problem.
  
  If, on the contrary, $u_s > |u_1|$, then the conformal factor $1/f$ is finite and smooth at
  all relevant values of $u$, and the situation precisely repeats the one in Penney's 
  solution in its branch $[3+]$, hence the system is catastrophically unstable.
     
  Lastly, in the intermediate case $u_s = |u_1|$, at the ``final point'' $u = u_s$,
  we have $g_{tt}^J$ tending to a finite limit, while $r_J \sim (u_s - u)^2 \to 0$
  (a center, where we have a curvature singularity, as is directly verified: for example, the    
  circumference to radius ratio of a small circle around it tends to infinity). Concerning \pbs,
  we find that the condition $|\df|/\phi < \infty$ leads to $|X|/(u_s - u) < \infty$. 
  Furthermore, since now $r \sim u_s - u$ and $z \sim (u_s - u)^3$, we obtain the condition
  $|Y|/z^{2/3} < \infty$ near the singularity. In the asymptotic solution \rf{Y3} this selects   
  the case $C_7 =0$. We thus have the boundary conditions  $|Y|< \infty$ as $z\to \infty$ and 
  $|Y|/z^{2/3} < \infty$ as $z \to 0$ for \eqn{Schr}.
      
\paragraph{Schwinger's theory \rf{Swg}.}  We now have 
\beq        \label{sol-Swg}
			f(\phi) = \phi = \frac K{(\psi-\psi_0)^2}, \qq
			ds_J^2 = \frac{(\psi-\psi_0)^2}{K}\, ds_E^2, \qq K > 0,\  \psi_0 = \const. 
\eeq
  The solution is \asflat\ at $u\to 0$ (at large $z$) under the condition $\psi_0 \ne 0$, 
  and behaves quite differently in the cases $\psi_0 >0$ and  $\psi_0 < 0$. 
  
  Since we have assumed $C >0$, in the case $\psi_0 > 0$, for branches $[1+]$ and $[2+]$
  with $u_1 > 0$, the range of $u$ is $u \in (0, u_s)$, where $u_s = \psi_0/C$, and the 
  singularity at $u = u_s$ is quite similar to that in Barker's theory  at $u = u_s$. For \pbs\ 
  we come again to the boundary condition  $|X|/(u_s- u) < \infty$ as $u \to u_s$ or, 
  equivalently, $|Y|/(z-z_s) < \infty$ as $z \to z_s$.
  
  In branch [$3+$] or $[1+]$ and $[2+]$  with $u_1 < 0$, we have an interplay of the 
  parameters $u_s$ and $u_1$ quite similar to the same branches in Barker's solution, 
  with the same behavior of background geometries and the same stability conclusions.
  
  In the case $\psi_0 < 0$, since now $\phi > 0$ everywhere, the solution is defined in the
  same range as Penney's in its all three branches, and its geometry is also almost
  similar to Penney's. 
  More precisely,  in branch [1+] with $u_1 > 0$, the conformal factor $1/f \sim u^2$
  at large $u$ is insignificant as compared to the exponential behavior of $\sinh[h(u+u_1)]$
  and $\sinh (ku)$. In other branches, where $u_{\max} < \infty$, the factor $1/f$, being 
  finite, also does not affect the qualitative properties of the metric. The only exception is 
  [2+] with $u_1 > 0$, in which case both $1/f$ and $s(h, u+u_1) \sim u^2$ at large $u$,
  the metric coefficient  $g^J_{tt}$ tends to a finite limit as $u \to \infty$.
  
  For \pbs\ we find that $\df/\phi = -2\dpsi/(\psi - \psi_0)$.
  Therefore, for branches $[1+]$ and $[2+]$ with $u_1 > 0$, where $\psi\to\infty$ at large 
  $u$, the condition $|\df/\phi| < \infty$ leads to $|\dpsi/\psi| < \infty$, similarly to Penney's 
  solution, with its hard instability. For branch $[3+]$ or $[1+]$ and $[2+]$  with $u_1 < 0$, 
  where $\psi$ has a finite limit as $u \to u_s$, the same condition leads to $|\dpsi| < \infty$, 
  again as for Penney's solution, and with the same result.  
  
  Thus in the solution with $\psi_0 < 0$ both the geometry (in almost all cases) and the 
  stability properties are similar to those of Penney's solution.    
  
\paragraph{Scalar fields nonminimally coupled to gravity, $f(\phi) = 1 - \xi \phi^2$.} The 
  particular form of solutions in the theories \rf{nonmin} and their geometric properties 
  depend on the value of the nonminimal coupling constant $\xi$. The following four cases 
  are distinguished:
\begin{description}   
\item[(i)]
	$\xi = 1/6$, conformal coupling.
	From \rf{phi-psi} we obtain, assuming $\phi^2 < 6$ (to provide $f >0$),  
\beq
		\psi-\psi_0 = \frac{\sqrt 3}{2} \log \frac{\sqrt 6 + \phi}{\sqrt 6  \phi}
		\ \ \then \ \
		\phi = \sqrt 6 \tanh \frac{\psi-\psi_0}{\sqrt 3},\ \ \ 
		 ds^2_J = \cosh^2 \frac{\psi-\psi_0}{\sqrt 3} ds_E^2.	
\eeq     

    In branch $[1+]$, $u_1 >0$, the behavior of this metric as ($u \to \infty$) depends on the 
    ratio $k/h$ or (related) $|C|/h$ (recall that here $k^2= h^2 + C^2$). Thus, if 
    $k < 2h \then |C| < h\sqrt 3$, then at large $u$  we have $r_J \to \infty$ and 
    $g^J_{tt}\to 0$, an attracting singularity at an infinitely growing sphere. On the contrary, 
    if $k > 2h \then |C| > h\sqrt 3$, we obtain a repulsive singular center, $r_J \to 0$ and 
    $g^J_{tt}\to \infty$, a \RN-like behavior. In the intermediate case $k = 2h$  and
    $|C| = h\sqrt 3$, the sphere $u=\infty$ is regular, and a conformal continuation leads 
    further either to a singular center ($r_J \to 0$), or to a traversable wormhole, or to an 
    extremal black hole. This continuation has been repeatedly studied, see, e.g,, 
    \cite{bbm70, br73, br-book, turok92, bar-vis2, stepan1},, 
    including, in particular, a stability analysis (though more in the uncharged case, $q=0$). 
    In the present study we assume $k \ne 2h$, so that $u =\infty$ is a singularity.     
   
\qq  At large $u$ we have 
    $f  = 1/\cosh^2 [(\psi-\psi_0)/\sqrt 3] \sim \e^{- 2\psi/\sqrt 3}$.
    Then, for perturbations, similarly to the Brans-Dicke theory, we have 
    $\delta f/f \sim \dpsi$, and the boundary condition as large $u$ again reads $|\dpsi| < \infty$.     
    
\qq    The same boundary condition is obtained for branch $[2+]$ with $h=0$ and $u_1 > 0$, 
    in which case the qualitative behavior of the metric $ds_J^2$ is the same as that of  $[1+]$
    at small $h$.
    
\qq    Lastly, in branch $[3+]$ as well as in $[2+]$ with $u_1 < 0$, the range of $u$ terminates 
    at finite $u=|u_1|$ just as in Penney's solution, with the same instability result.
	
\item[(ii)]
	$0 < \xi < 1/6$. According to \rf{phi-psi}, the fields $\phi$ and $\psi$ are related by
\beq                        \label{xi-2} 
		\sqrt 2 d\psi = \frac {\sqrt{1 - \eta \phi^2}}{1 -\xi\phi^2}d\phi,
		\qq
		\eta = \xi(1 - 6\xi); \qq \eta >0.
\eeq	
	Integrating, we obtain, in agreement with \cite{bar-vis2, stepan1}, 
\bearr                       \label{psi-2} 
		\psi - \psi_0 = \frac{\sqrt 3}{2} \log \Big[B(\phi) H^2(\phi)\Big],
\\ \lal                       \label{B-2} 
		B(\phi) = \frac{\sqrt{1-\eta\phi^2} + \sqrt 6 \xi\phi}{\sqrt{1-\eta\phi^2} - \sqrt 6 \xi\phi},
\\ \lal                       \label{H-2} 
           \log H(\phi) = \frac{\sqrt\eta}{\sqrt 6 \xi} \arcsin(\sqrt\eta \phi).
\ear			 
     It is easy to see that $\phi \to 1/\sqrt\xi$ leads to $\psi \to \infty$, which in turn corresponds 
     to $u \to \infty$. The function $H(\phi)$ is finite and smooth in the whole range 
     $\phi^2\leq 1/\xi$, therefore, the qualitative properties of the solution in \MJ\ are completely
     determined by $B(\phi)$, and the results are quite similar to those for $\xi = 1/6$. In particular, 
     the solution with $k = 2h$ (belonging to branch  $[1+]$) requires a conformal continuation 
     and is excluded from the present analysis. At $k \ne 2h$, at large $\psi$ the function 
     $f(\phi(\psi))$ behaves as $\e^{-2\psi/\sqrt 3}$, and for perturbations we obtain the same
     boundary condition as $u \to \infty$: $|\dpsi| < \infty$. 
     
\qq  For branches  $[2+]$ and $[3+]$, both the qualitative features of the solution and the 
     boundary conditions for \pbs\ are the same as for $\xi = 1/6$.
	
\item[(iii)]
	$\xi > 1/6$. In this case, \eqs \rf{xi-2}--\rf{B-2} are again valid, though now $\eta < 0$,
	which results in another expression for the function $H(\phi)$:
\beq                       \label{H-3} 
			\log H(\phi) = -\frac{\sqrt{-\eta}}{\sqrt 6 \xi} \sinh^{-1} (\sqrt{-\eta} \phi).
\eeq	 
     Still, quite similarly to $\xi < 1/6$, this $H(\phi)$ is finite and smooth at all $\phi^2 < 1/\xi$,
     and therefore the whole further reasoning simply repeats that for $0 <\xi <1/6$, with the
     same results.

\item[(iv)]
	$\xi < 0$. 	Equations \rf{xi-2}--\rf{B-2} are again valid, while now both $\xi < 0$ and 
	$\eta < 0$. Equation \rf{H-3} turns out to be also valid again, but, unlike the previous cases, 
	here the $\phi$ field is defined for all $\phi \in \R$, At large $|\phi|$ we always have 
	$\psi - \psi_0 \sim \log |\phi|$, therefore, in the cases where $\psi \to \infty$
	(that is, for branches $[1+]$ and $[2+]$ with $u_1 > 0$),
	we have $\delta\psi \sim \df/|\phi|$, hence $\delta f/f \sim \dpsi$, and consequently 
	the boundary condition for \pbs\ reads $|\dpsi| < \infty$. One can also notice that 
     no conformal continuations are observed in this case.
     
\qq	The solutions of branches $[3+]$ and $[2+]$ with $u_1 <0$, defined up to a finite 
	value of $u$, behave in the same way as with $\xi >0$.
\end{description}  

\subsection{Boundary-value problems}

  We have seen that in all enumerated theories there are branches of the solutions exhibiting
  the same catastrophic instabilities as Penney's solution. However, there are others that
  require a further study. In all of them, the effective potential $\Veff$ has the same form 
  \rf{Veff-u} for all solutions under study, but it is expressed in terms of $u$ instead of 
  the tortoise coordinate $z$ used in \eqn{Schr}, while the boundary conditions are equally 
  well formulated using $z$ or $u$. 
  
  We have used the asymptotic form of $\Veff$ at large and small $z$ to obtain the
  asymptotic form of possible solutions to \eqn{Schr} and further to obtain certain instability 
  conclusions. Also, the asymptotic form of $\Veff$ at large $z$ (at spatial infinity), 
  namely, $\Veff \approx 2m/z^3$, indicates that $\Veff >0$ at sufficiently large $z$,
  which means that our solutions must be stable if defined only for such $z$, or, in other 
  words, only at sufficiently small $u$. It can happen in those theories where $\psi = Cu$ is 
  bounded above (like, e.g., Barker's theory), which in turn bounds the values of $u$. 
  
  However, for a further numerical investigation in 
  the remaining cases, it appears to be be better to use the master equation \rf{eq-X}
  in terms of our harmonic coordinate $u$, with which all coefficients are known explicitly.  
  In terms of $u$, \eqn {eq-X} has the form
\beq  			\label{eq-XX}
			X'' + [r^4(u) \Omega^2 - W(u)]X =0, \qq
			W(u) = \frac {2 C^2\e^{2\alpha(u)}}{r'{}^2}\bigg(\frac{q^2}{r^2}-1\bigg),
\eeq  
  where $r(u) = \e^\beta$ and $\e^{2\alpha(u)} = - g^E_{11}$ are taken from the metric
  \rf{ds_E} in \ME. 
  
  The asymptotic form of solutions to \eqn{eq-XX} at the ends of the range of $u$
  must correspond to that of \eqn{Schr} in terms of $z$ obtained above. Let us verify that. 
  
  First, as $u\to 0$ ($r \approx 1/u \to \infty$), $W \approx -2C^2$, and for $\Omega^2<0$
  \eqn{eq-XX} takes the form 
\beq
		X'' - |\Omega|^2 X/u^4 = 0 \ \ \then \ \ 
		X = u\big(C_1\e^{|\Omega|/u} + C_2\e^{-|\Omega|/u}\big), 
\eeq
   and this solution precisely corresponds to \rf{Y0+}. For $\Omega =0$, \eqn{eq-XX}
   at small $u$ reduces to $X'' + 2C^2 X =0$, and its evident solution completely 
   corresponds to \rf{Y00}. In both cases, the boundary condition $X\to 0$ as $u\to 0$ 
   selects one of the two  linearly independent solutions.
   
   Second, in the cases where  $u \to \infty$, \eqn{eq-XX} takes the approximate form
\beq                    			\label{eq-X1}  \wide
   	   X'' + \big[\Omega^2 \e^{4(h-k)u} - W_0^2 \e^{-2hu} \big]X =0,
\eeq   
   where $W_0 = \const >0$, and $k > h \geq 0$. The expression in brackets is 
   dominated by the first term if $k < 3h/2$, by the second term if $k > 3h/2$, and 
   both are of equal significance if $2k=3h$ (recall that now $k^2 = h^2 + C^2$). 
   In any case, if $\Omega^2 \leq 0$, this equation in the leading approximation reads
\beq             			\label{eq-X2}
   		X'' - K^2 \e^{-2nu} X =0 
\eeq   		
   with some positive constants $K$ and $n$, and its solution has the form 
   $X = Z_0\big((K/n)\e^{-nu}\big)$, where 
   $Z_0(x) = C_J J_0(x) + C_N N_0(x)$ is a solution of the Bessel equation with zero 
   index, $J_0$ and $N_0$ being the Bessel and Neumann functions, respectively.  
   Since we are considering $u\to \infty \ \then \e^{-nu}\to 0$, we obtain in this limit, 
   according to the properties of the Bessel and Neumann functions at small 
   values of their argument,
\beq                              
		X \approx C_3 + C_4 u,
\eeq   
   in full agreement with \rf{Y1,Y2}, and a possible requirement $|X| < \infty$ selects
   solutions with $C_4 =0$.
   
   Third, in the cases where  $u \to u_s$ such that $s(h(u+u_1))\to 0$, that is, near
   RN-like singularities of Penney's solution, \eqn{eq-XX} takes the approximate form
   $X'' - W_0 X =0$ (the term with $\Omega^2$ is negligible since it contains a
   factor $\sim (u_s-u)^4$), and $u=u_s$ is a regular point of this equation, where
   $\psi = Cu$ is finite, so the function $f(\psi)$ of our STT is generically also regular
   and finite,  leading to the natural requirement $|X| \sim |\dpsi|< \infty$.
   As we have seen, in such cases we obtain the same hard instability as in Penney's 
   solution. 

   Fourth, if the conformal factor $1/f \to 0$ at some $u = u_s < \infty$ at which the 
   metric \rf{ds_E} is regular, it makes the STT solution terminate at this value of $u$,
   and we must formulate the boundary condition there. In our examples, we have at such 
   points $f \sim (u_s - u)^{-2}$, hence $\delta f/f \sim \dpsi/(u_s - u)$ (since $\psi = Cu$),
   leading to the boundary condition 
\beq
		|X|/(u_s - u)  < \infty,
\eeq   
   in agreement with our inferences for particular STT.
   
   We can summarize that the cases where a further numerical study is necessary reduce
   to the following boundary-value problems:
    
\begin{description}
\item[\bf Problem 1.]
		Equation \rf{eq-XX}, range $u\in (0, u_s)$, boundary conditions: 
		$X \to 0$ as $u \to 0$, and $|X|/(u_s - u) < \infty$ as $u \to u_s$.   	
\item[\bf Problem 2.]
		Equation \rf{eq-XX}, range $u\in (0,\infty)$, boundary conditions: 
		$X \to 0$ as $u \to 0$, and $|X| < \infty$ as $u \to \infty$. 
\end{description}  
  
   By solving these problems, we try to make clear whether or not there exist physically 
   meaningful \pbs\ (i.e., those satisfying the above boundary conditions) corresponding to
   eigenvalues $\Omega^2 \leq 0$, and this would mean that the static 
   background configuration is linearly unstable. 
  
\section{Numerical analysis}  
\subsection{Preliminaries}
   
   The boundary-value problems have now been posed; meanwhile,
   we notice that Problem 1 must be solved with five different analytical forms of 
   the potential $W(u)$ corresponding to all three branches of the background solution
  (the third branch, in turn, contains solutions with different signs of the parameter $k$),
   while Problem 2 deals with only two forms of $W(u)$ for the branches where $u \in (0, \infty)$. 
   Without loss of generality we will put the scalar charge parameter $C$, which is nonzero 
   in all solutions under consideration, to be $C=1$, thus actually fixing the length scale.  
   
\paragraph{Problem 1} can be solved by the ``shooting'' method starting from $u = u_s$: we 
   numerically solve \eqn{eq-XX} with different $\Omega^2 \leq 0$ in the range 
   $u\in (0, u_s)$ with the boundary conditions  
\beq   \label{BC1}
		   X(u_s) =0, \qq  X'(u_s) = 1
\eeq		   
   and select the values of $\Omega$ for which the solution yields $X(0) =0$.
   (The particular value of $X'(u_s)$ is insignificant since \eqn{eq-XX} is linear.)
   The function $r(u)$ is, according to \rf{ds_E}, 
\beq
			r(u) = \frac {|q| s(h, u+u_1)}{s(k,u)},
\eeq   
   and the potential function $W(u) $ is given in \rf{eq-XX}, or more specifically,
\beq                           \label{W(u)}
   	W(u) = \frac{2C^2}{s^2(k,u) \Big[\dfrac{s'}{s}[h, u+u_1]-\dfrac{s'}{s}(k,u)\Big]^2}
   				\bigg[\frac{s^2(k,u)}{s^2[h(u+u_1)]} -1 \bigg],
\eeq
   where the functions ``$s()$'' are defined in \eqs \rf{beta} and \rf{gamma}; 
   and the corresponding expressions $s'/s$ are found accordingly, for example, 
   $\frac{s'}{s}(k,u) = k \coth (ku)$ for $k >0$, and so on. The calculations must be
   carried out separately for the cases (1) $k > h >0$ (branch $[1+]$), (2) $k > h =0$
   (branch $[2+]$), (3) $k > 0, h <0$, (4) $k =0, h <0$, and (5) $h < k <0$
   (branch $[3+]$ in the last three cases).
  
\paragraph{Problem 2} is more complicated because of the infinitely remote ``right end'' 
   $u \to \infty$. In practice, to perform calculations, we have to move the boundary to 
   some large but finite value $u_0$ and specify the values of $X(u_0)$ and $X'(u_0)$
   corresponding to a solution admitted by our boundary conditions. 
   
   Thus, for branch $[1+]$ ($k>h>0$), \eqn{eq-XX} at large $u$ takes the form \rf{eq-X2}. 
   However, the leading approximation $X(\infty) = X_0$ is insufficient for specifying the 
   boundary conditions. Solving \eqn{eq-X2} in the next approximation in powers
   of $\e^{-nu}$, we obtain (putting $X(\infty) =1$ without loss of generality)
\beq         \label{X1}
		 X(u) \approx 1 - \frac{K^2}{4n^2} \e^{-2nu} \qq (u \to \infty),
\eeq      
   making it possible to approximately specify $X$ and $X'$ at some large $u = u_0$, able 
   to serve as boundary conditions for the shooting method:
\beq          \label{BC2a}
		X(u_0) = 1 - \frac{K^2}{4n^2} \e^{-2nu_0},\qq
		X'(u_0) = \frac{K^2}{2n} \e^{-2nu_0},		
\eeq   
   where the parameters $K$ and $n$ are determined from \eqn{eq-X1} as described above.   
   
   For Branch $[2+]$ ($k>h=0$), in \eqn{eq-XX} the term $\Omega^2 r^4$ is insignificant 
   at large $u$ because $r(u)$ decays exponentially as $u \to \infty$, while $W$ shows a power-law  
   decrease. In the present case, we have $u_1 = 1/q >0$ and $k =|C|$, and the 
   approximate form of \eqn{eq-XX} at large $u$ reads
\beq
			X'' - \frac{2 X}{(u + u_2)^2} = 0, \qq    u_2 = u_1 - 1/k,
\eeq      
   and is solved by 
\beq       \label{X2}
                  X \approx C_9 (u + u_2)^{-1} + C_{10} (u + u_2)^2.
\eeq   
   The requirement $X < \infty$ makes us put $C_{10}=0$. Under this condition,
   the expression \rf{X2} allows us to specify $X$ and $X'$ at some large $u = u_0$,
   to be used as boundary conditions for \eqn{eq-XX}:
\beq          \label{BC2b}
		X(u_0) = \frac{1}{u_0 + u_2},\qq
		X'(u_0) = -\frac{1}{(u_0 + u_2)^2}.
\eeq      
   
\subsection{Boundary-value problem 1   }

   Now we try to solve \eqn{eq-XX},
\beq                      \label{eq-XA}
			X'' + [r^4 (u) \Omega^2 - W(u)] X =0, 
\eeq
   by the shooting method described above for different values of the solution parameters, 
   including $u_s$, and different $\Omega^2$.  The independent parameters of the problem 
   are those of the metric in \ME\ that can be chosen as the scalar charge $C$, the electric
   charge $q$, and $h$ related to the mass $m$ according to \rf {m}, while the constants $k$ and 
   $u_1$ are connected with $C, q, h$ by the relations \rf{int} and \rf{u_1}, that is,  
\beq
		k^2 \sign k = C^2 + h^2 \sign h, \qq 		s^2 (h, u_1) = 1/q^2,
\eeq   
   the latter following from $g_{00} = 1$ at $u=0$ (spatial infinity). One more parameter, $u_s$, 
   emerges at a transition to \MJ\ in scalar-tensor theories.
   
   We apply the standard procedure of ODE solving, which yields a numerical curve 
   $X_{\rm num}(u)$ corresponding to a chosen test value of $\Omega^2$. When this value is 
   close to an eigenvalue of the problem with \eq \eqref{eq-XA}, the sign of the curve 
   $X_{\rm num}(u)$ strongly fluctuates near $u=0$. Tracking its behavior, one can determine 
   such a critical value of $\Omega^2$ as a candidate eigenvalue with necessary accuracy.

\paragraph{Branch [1+]:} $k > h >0$, and $k^2 = h^2 + C^2$.
   In \eqn{eq-XA} we now have
\bearr                        \label{W1+}
		u \in (0, u_s), \qq     u_s < |u_1| \text{ for }  u_1 < 0; \qq 
		r(u) = \frac {|q| k \sinh [h(u+u_1)]}{h \sinh (ku)},
\nnn
		W(u) = \frac{2 C^2} {\Big[ k \coth(ku) - h \coth [h (u+u_1)]\Big]^2}
   				\bigg[\frac{h^2} {\sinh^2[h(u+u_1)]} - \frac{k^2}{\sinh^2(k,u)}\bigg],				
\ear   
   with the boundary conditions: $X(u_s) = 0, X'(u_s) =1$ (a position for shooting) and 
   $X(0) =0$ (as a target).

    Our numerical analysis shows that there are instability regions for $u_1 > 0$. As to $u_1<0$, 
    we did not find any instability regions for a sufficiently wide range of parameters.

\begin{figure}[H]
\centering
\includegraphics[width=7.5cm]{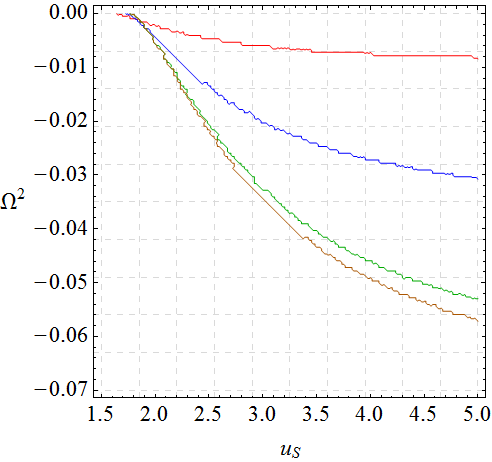}\qq
\includegraphics[width=7.5cm]{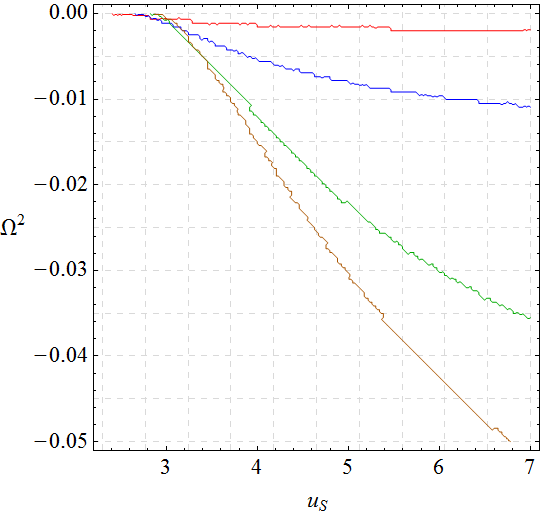}
\caption{\small 
	Boundary-value problem 1, branch [1+]. The eigenvalues $\Omega^2<0$ depending on the 
	choice of $u_s$ for some values of $h$ and $u_1$. Each curve is in fact a separatrix between 
	numerically obtained regions of $X_{\rm num}(u) > 0$ as $u\to 0$ (above the line) from 
	regions of $X_{\rm num}(u)<0$ (below the line). \underline{Left}: Case $h=1$, and 
	$u_1=0.5, 1, 2, 3$ upside-down.  \underline{Right}: Case $h=0.5$, and $u_1=0.5, 1, 2, 3$ 
	upside-down.}
	\label{FigOmega1}
\end{figure}

\begin{figure}[ht]
\centering
\includegraphics[width=7.5cm]{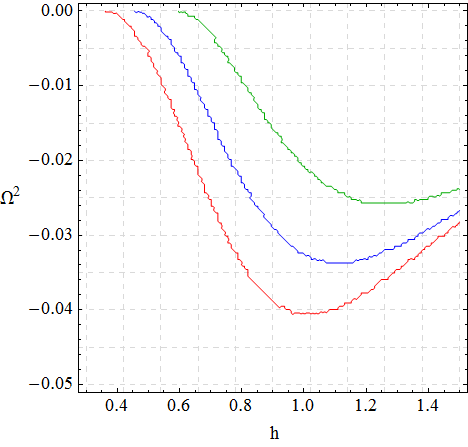}\qq
\includegraphics[width=7.5cm]{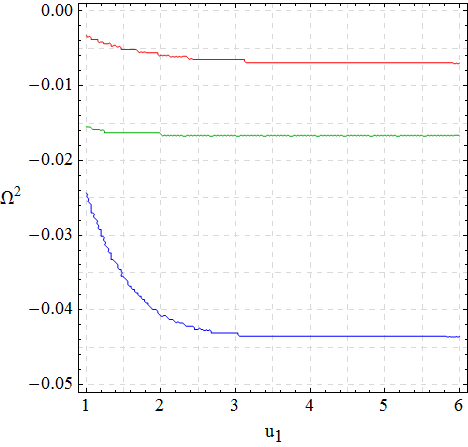}
\caption{\small 
	Boundary-value problem 1, branch [1+]. \underline{Left}: The eigenvalue $\Omega^2<0$ as a 
	function of $h$ for $u_1=2$, $u_s = 2.5, 3, 3.5$, upside-down. \underline{Right}:  The eigenvalue 
	$\Omega^2<0$ as a function of $u_1$ for $u_s = 3.5$, $h = 0.5, 1, 2$, upside-down.}
	\label{FigOmega2}
\end{figure}

    The results of our stability analysis for branch [1+] are depicted in Fig.\,\ref{FigOmega1}. 
    (possible eigenvalues $\Omega^2<0$ are shown as a function of $u_s$ for certain values of 
    $h$ and $u_1$) , while Fig.\,\ref{FigOmega2} shows $\Omega^2$ as a function of $h$ for 
    some values of $u_1$ and $u_s$, and as a function of $u_1$ for some values of $h$ 
    and $u_s$). In this and other figures, a slight jaggedness of the curves is an artifact  of the
    numerical procedure used for constructing the maps of the existing values of $\Omega^2$
    using the shooting method. 

   One can see that the instability (i.e., the existence of $\Omega^2<0$) disappears at sufficiently 
   small $h$ (Fig.\,\ref{FigOmega2}, left panel) or sufficiently small $u_s$,
   that is, an instability takes place for $u_s \geq u_{s, \rm crit}$, where the critical value 
   $u_{s, \rm crit}$ depends on $h$ and $u_1$ (Fig.\,\ref{FigOmega1}). 
   Also, at large $u_1$ we obtain an asymptotic value of $\Omega^2$ (Fig.\,\ref{FigOmega2}, 
   right panel) corresponding to the case of zero charge, thus approaching the results of 
   our previous paper \cite{we23}.
   
\begin{figure}[ht]
\centering
\includegraphics[width=10cm]{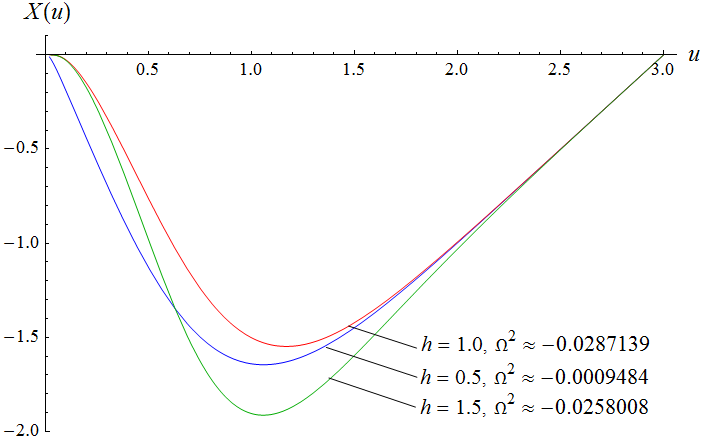}
\caption{\small 
	Boundary-value problem 1, branch [1+]. Examples of numerical curves $X(u)$ with some 
	eigenvalues $\Omega^2$ for different $h$ and the fixed values $u_1=1.5, u_s=3$.}
	\label{FigOmega3}
\end{figure}

   Examples of numerical solutions $X(u)$ with some of the eigenvalues $\Omega^2$ 
   are shown in Fig.\,\ref{FigOmega3}.

\paragraph{Branch [2+]:} $k > h =0$, and $k^2 = C^2$.  
   We now have
\bearr                         \label{W2+}
	u \in (0, u_s), \qq u_s < |u_1| \ \text{only for}\  u_1 < 0; \qq  r(u) = \frac {|q| k (u+u_1)}{\sinh (ku)},
\nnn
		W(u) = \frac{2 k^2} {\Big[ k \coth(ku) - 1/(u+u_1) \Big]^2}
   				\bigg[\frac{1}{(u+u_1)^2} - \frac{k^2}{\sinh^2 (ku)} \bigg],				
\ear   
   with the boundary conditions: $X(u_s) = 0, X'(u_s) =1$ (a position for shooting) and 
   $X(0) =0$ (as a target).   
   
   A numerical analysis did not reveal any instability regions in this branch 
   for a sufficiently wide range of the parameters.
   
\paragraph{Branch [3+]:} $h < 0$,  and the solution behavior is mostly governed by the 
   function $\sin [|h|(u+u_1)]$. Now, without loss of generality,
\beq   
		   u_1 \in (0, \pi/|h|), \ \ \  {\rm and} \ \ \ u \in (0, u_s), \ \ \    u_s < u_1.
\eeq
   As before, the boundary conditions are: $X(u_s) = 0, X'(u_s) =1$ (a position for shooting) and 
   $X(0) =0$ (as a target). However, for $r(u)$ and $W(u)$ we have different analytical expression
   in the following three subcases: 
\besub   
\bearr \nq
		{\bf [3{+}a]:} \  k>0, \  \ k^2 = C^2 - h^2, \qq 
		  r(u) = \frac {|q| k \sin [|h|(u+u_1)]}{|h| \sinh (ku)},
\nnnv  \qq
		W(u) = \frac{2 C^2} {\Big[ k \coth(ku) - |h| \cot [|h| (u+u_1)]\Big]^2}
   				\bigg[\frac{h^2} {\sin^2[|h|(u+u_1)]} - \frac{k^2}{\sinh^2(k,u)}\bigg],				
\yyy   \nq
		{\bf [3{+}b]:} \  k = 0, \  \  h^2 = C^2, \qq  	r(u) = \frac {|q| \sin [|h|(u+u_1)]}{|h|u},
\nnn  \qq
		W(u) = \frac{2 C^2} {\Big[ 1/u - |h| \cot [|h| (u+u_1)]\Big]^2}
   				\bigg[\frac{h^2} {\sin^2[|h|(u+u_1)]} - \frac{1}{u^2}\bigg],				
\yyy   \nq
		{\bf [3{+}c]:}  \ h < k < 0, \ \ k^2 = h^2 - C^2, \qq	
					r(u) = \frac {|q| |k| \sin [|h|(u+u_1)]}{|h| \sin (|k|u)},
\nnn \qq
		W(u) = \frac{2 C^2} {\Big[ |k| \cot(|k|u) - |h| \cot [|h| (u+u_1)]\Big]^2}
   				\bigg[\frac{h^2} {\sin^2[|h|(u+u_1)]} - \frac{k^2}{\sin^2(|k|u)}\bigg],				
\ear     
\esub
   
   Again, in our numerical study, we did not find any instability regions in this branch for a 
   sufficiently wide range of the parameters.
   
\subsection{Boundary-value problem 2   }

   We now again try to solve \eqn{eq-XA} by the shooting method with different $\Omega^2$, 
   but now for the branches of Penney's solution with $u \in \R_+$, therefore, instead of $u_s$,
   we must choose a large enough value $u_0$ as a shooting position with the boundary values 
   of $X$ and $X'$ specified by \rf{BC2a} and \rf{BC2b}. More specifically:

\paragraph{Branch [1+]:} $k > h >0$, $u_1 > 0$, and $k^2 = h^2 + C^2$; the quantities $r^2(u)$
   and $W(u)$ are given in \rf{W1+}. 
   The boundary conditions at $u=u_0$ for \eqn{eq-XA} have the form \rf{BC2a}, that is:
\besub   
\bearr   
        X(u_0) = 1 - \frac{K^2}{4n^2} \e^{-2nu_0},\qq
		X'(u_0) = \frac{K^2}{2n} \e^{-2nu_0},		
\nnnv   \tall
	k < 3h/2: \qq   K^2 = |\Omega^2|, \qq     n = 2 (k-h);
\yyy   
	k = 3h/2: \qq   K^2 =  |\Omega^2| + W_0^2, \qq  n = h; 	
\yyy     
	k > 3h/2: \qq   K^2 = W_0^2,  \qq n = h,
\qq
			W_0^2 = \frac{8 h^2 (k+h)}{k^2 (k-h)} \e^{-2 h u_1}. 		
\ear   
\esub

\begin{figure}[ht]
\centering
\includegraphics[width=7cm]{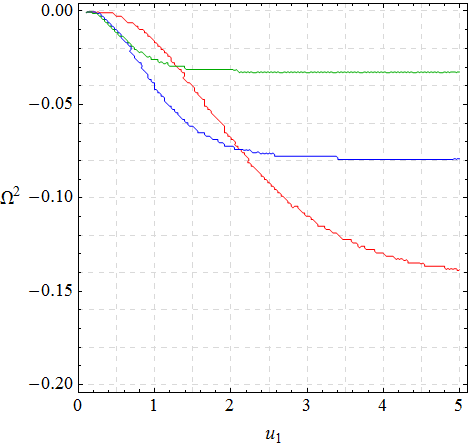}\qq
\includegraphics[width=7cm]{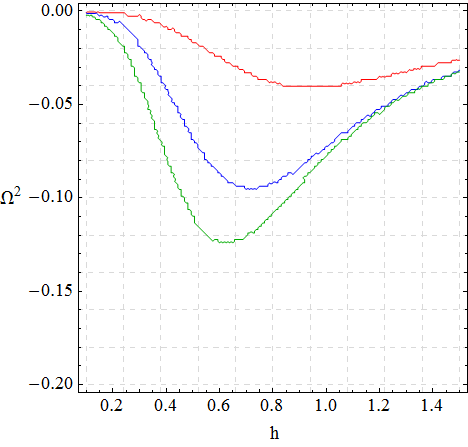}
\caption{\small
	Boundary-value problem 2, branch [1+]. \underline{Left}: The eigenvalue $\Omega^2<0$
	as a function of $u_1$ for $h = 0.5, 1, 1.5$, bottom-up. \underline{Right}:  
	The eigenvalue $\Omega^2<0$ as a function of $h$ for $u_1 = 1, 2, 3$, upside-down.}
	\label{FigOmega4}
\end{figure}

   The results of our numerical analysis for this branch are shown in Fig.\,\ref{FigOmega4}. The 
   diagrams indicate instability regions with $\Omega^2<0$ for a certain set of the parameters 
   $h$ and $u_1$. One can see that decreasing these parameters corresponds to the limit 
   $\Omega^2\to -0$. Similarly to Problem 1, at given $u_1$ there is a range of $h$ such that the 
   instability takes place at $h \geq h_{\rm crit}(u_1)$. 
   
\begin{figure}[ht]
\centering
\includegraphics[width=9.5cm]{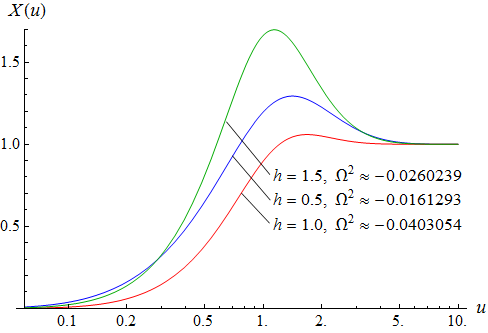}
\caption{\small 
	Boundary-value problem 2, branch [1+]. Examples of numerical solutions $X(u)$ with some 
	eigenvalues $\Omega^2$ for different $h$ and the fixed value $u_1=1$.}
	\label{FigOmega5}
\end{figure}

   Figure \ref{FigOmega5} shows examples of numerical solutions $X(u)$ with some of the 
   eigenvalues $\Omega^2$ for different values of $h$.

\begin{figure}[ht]
\centering
\includegraphics[width=9.5cm]{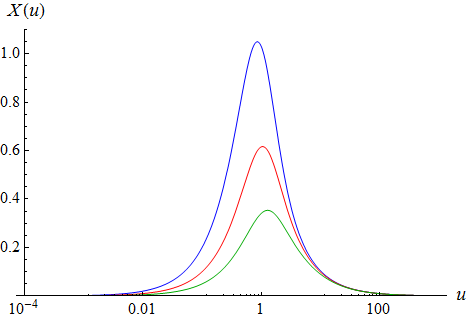}
\caption{\small
	Boundary-value problem 2, branch [2+]. Examples of numerical solutions $X(u)$ 
	corresponding to the eigenvalue $\Omega^2=0$ for $u_1=0.5, 1, 2$ upside-down.}
	\label{FigOmega6}
\end{figure}

\paragraph{Branch [2+]:} $k > h =0$, $u_1 > 0$, and $k^2 = C^2$; the quantities $r^2(u)$
   and $W(u)$ are given in \rf{W2+}. 
   The boundary conditions at $u=u_0$ for \eqn{eq-XA} have the form \rf{BC2b}, that is,
\beq          \label{BC2+}
		X(u_0) = \frac{1}{u_0 + u_2},\qq
		X'(u_0) = -\frac{1}{(u_0 + u_2)^2}, \qq
		u_2 = u_1 - 1/k.
\eeq      
  In all cases, the ``target'' is $X(0) =0$.
  
  A numerical study did not reveal any nontrivial instability regions in this branch for a 
  sufficiently wide range of parameters, except for the case $\Omega^2=0$, which corresponds 
  to a numerical solution $X_{\rm num}(u)$ with the proper behavior at $u\to 0$ with 
  necessary accuracy. Examples of such numerical solutions $X(u)$ for different $u_1$ are 
  shown in Fig.~\ref{FigOmega6}

  The results of solving Problems 1 and 2 are summarized in Table 1. 
   
\begin{table*}[h]
\caption{Stability results from solving boundary-value problems (BVP) 1 and 2}
\begin{center}
\small
\begin{tabular}{|c|c|c|l|}    
\hline  
    Problem             & Range   &   Solution                                      & Results$^*$  \tall
\\[2pt]  \hline  \tall
		     & \raisebox{-3mm}{$u \in (0, u_s)$,} &  [1+], $u_1 < 0$, $u_s  < |u_1|$    & stable
\\[-2pt]
	\ BVP 1  & \raisebox{-3mm}{$u_s < \infty$} &  [1+], $u_1 > 0$   
										& unstable if $u_s \geq u_{s, \rm crit}$
\\[-2pt] 
	              &  &  [2+], [3+]  			& stable
\\[2pt] \hline  \tall
    \ \raisebox{-3mm}{BVP 2}    & \raisebox{-3mm}{$u \in (0, \infty)$} &  [1+], $u_1 > 0$   
    										& unstable if $h \geq h_{\rm crit}$
\\[-2pt]
	              &  &  [2+], $u_1 > 0$           & unstable with $\Omega =0$	
\\[2pt] \hline
\end{tabular}   
\end{center}   
  ${}^*$ The word ``stable'' means that solutions with $\Omega^2 \leq 0$ have not been found; 
  the quantities	$u_{s, \rm crit}$ and $ h_{\rm crit}$ are finite critical values of the corresponding
  parameters.
\end{table*}                
\begin{table*}[ht]
\caption{Electrovacuum STT solutions: Stability under monopole perturbations}
\begin{center}
\small  
\begin{tabular}{|p{65mm}|p{77mm}| l |}
\hline
	\cm Theory, $f(\phi(\psi))$    & Solution branches, $u_{\max}$, type of singularity 
								   & Results$^{**}$   \tall
\\[2pt]  \hline  \tall
		GR, $\phi = \psi$, $f \equiv 1$   &
								[1+], [2+], $u_1>0$; $u \to\infty$, scalar type$^{***}$ & unstable+
\\[2pt]
		\cm (Penney's solution)   &  
								[1+], [2+], $u_1 < 0$, or [3+];  $u\to |u_1|$, RN type.  &   unstable+
\\[2pt]  \hline  \tall
		Brans-Dicke, $\omega>-\frac 32$; at $\omega=0$, $k \ne 2h$  &   
								[1+], [2+], $u_1 >0$; $u \to\infty$ 				& BVP 2
\\[2pt]
		\cm $f = \phi = \exp\Big[ \frac{\sqrt{2} (\psi-\psi_0)}{\sqrt{\omega+3/2}}\Big]$    &
							[1+], [2+], $u_1 < 0$, or [3+];  $u\to |u_1|$, RN type  &   unstable+
\\[2pt] \hline  \tall
		Barker,           &   [1+], [2+], $u \to u_s =(\pi/2+\psi_0)/C <  |u_1|$ & BVP 1 
\\[2pt] 
		\cm $f = \phi = \dfrac 1{\cos^2(\psi-\psi_0)}$  &
						        [1+], [2+], $u_1 >0$; $u \to u_1 > u_s$, RN type   &   unstable+
\\[2pt] 
						&	[1+], [2+], $u_1 < 0$, or [3+];  $u\to |u_1|$, RN type  &   unstable+
\\[2pt] \hline  \tall
		Schwinger,    &	 $\psi_0 > 0$, [1+], [2+], $u \to u_s =\psi_0/C < |u_1|$ & BVP 1
\\[2pt] 
		\cm $ f= \phi = \dfrac K{(\psi-\psi_0)^2} $  &  $\psi_0 \geq 0$,
							 [1+], [2+], $u \to u_1 > u_s $, RN type   &   unstable+
\\[2pt] 
						  &  $\psi_0 < 0$, all branches similar to Penney's    	 & unstable+
\\[2pt] \hline    \tall
		Nonminimal coupling,  &
								[1+], [2+], $u_1 >0$; $u \to\infty$ 				& BVP 2
\\[2pt]  \cm
	   $\xi >0$, $k\ne 2h$, or $\xi < 0$  
					&	    [1+], [2+], $u_1 < 0$, or [3+];  $u\to |u_1|$, RN type  &   unstable+
\\[2pt] \hline
\end{tabular}    
\end{center}
\small
   	${}^*$ BVP = boundary-value problem.\\
   	${}^{**}$ ``unstable+'' means that a perturbation increment is indefinite, as it is for 
   		Penney's solution.\\
	${}^{***}$ A scalar type singularity is characterized by $\psi \to \infty$, $r \to 0$, and 
		$g_{tt} \to 0$.	  
\end{table*}                

\section{Results and discussion }

   We have considered the linear stability problem for space-times conformally related 
   to the one obtained by R. Penney \cite{penney} and comprising electrovacuum solutions for
   the Bergmann-Wagoner-Nordtvedt class of STTs of gravity with massless scalar fields. 
   Using the fact that the conformal mapping from Jordan (\MJ) to Einstein (\ME) manifold
   is simply a change of variables in the field equations, the perturbation equations in \MJ\
   are reduced to those in \ME\ which  are common to all STT of the  class under consideration
   and coincide with those of GR. However, the boundary conditions that select physically 
   meaningful perturbations, depend on a particular STT and even on the properties of 
   particular solutions. Thus, in many cases, the whole manifold \MJ\ maps to only a region 
   in \ME\ (particularly, in solutions where the variable $u$ is restricted to $u < u_s$ in \MJ\
   while its range is wider in \ME), and the boundary conditions for \pbs\ must be imposed on
   a regular intermediate surface from the viewpoint of \ME. 
   
   For some STT solutions, the boundary conditions coincide with those imposed in the 
   framework of \ME --- in such cases the stability conclusions obtained in \ME\ are naturally
   extended to \MJ. In other cases, it is necessary to consider other boundary-value
   problems. This work has been carried out here for some well-know examples of STT, and 
   its results are summarized in Table 2, with reference to the outcome of solving Problems 1 
   and 2 presented in Table 1. 

   A general observation that follows from the present study is a stabilizing role of the electric
   or magnetic charge in some STT (unlike GR), where whole branches of the electrovacuum 
   solutions turn out to be stable under radial \pbs. On the other hand, there are branches of 
   charged solutions with RN-type singularities that inherit the instability property of Penney's solution, 
   in the same manner as many of the STT extensions of Fisher's scalar-vacuum solution
   inherit its instability \cite{we23}. In fact, the results obtained in \cite{we23} are reproduced 
   from the present ones by considering the limit $q \to 0$ in branch [1+] of the solutions.
   Indeed, since in this branch (see \rf{gamma} and \rf{ds_E}) $q^2 = h^2 [\sinh (hu_1)]^{-2}$, 
   the limit $q\to 0$ at a fixed value of $h$ corresponds to $u_1 \to \infty$, and consequently 
   the expression for $\e^{2\gamma}$ in \rf{ds2} and \rf{ds_E} reduces in this limit to
   $\e^{2\gamma} = \e^{-2hu}$ that characterizes Fisher's solution; a further substitution 
   $\e^{-2ku} =  1- 2k/x$ leads to Fisher's metric in its well-known form \rf{ds+}
   with $\e^{2\gamma} = (1 - 2k/x)^{a},\ a = h/k$.

   For those solutions that turn out to be stable under monopole perturbations, a stability study 
   must in general be extended to higher multipolarities; however, as discussed in the introduction,
   most probably such a study will not reveal any instabilities, and this observation is confirmed by
   the previous stability studies, e.g., \cite{stab12, stab13}.  
      
   Quite evidently, the present approach and results can be extended to many more particular 
   STTs since the boundary-value problems 1 and 2 for \sph\ perturbations emerge under quite 
   general conditions. In particular, it is true for STT representations of various modified theories 
   of gravity, such as, for example, the hybrid metric-Palatini gravity \cite{hybrid}
   and its generalized versions \cite{g-hybrid}, for which \ssph\ solutions are studied in 
   \cite{we-hyb, we-g-hyb}, as well as different nonlocal and high-order theories of gravity, 
   see, e.g., \cite{od17-rev, od08, kb-eliz10, chrv22}.
   
   It should also be noted that the present results only apply to STT with massless scalar
   fields, while there are numerous theories with nonzero scalar field potentials $U(\phi)$,
   see \eqn{S_J}, and their solutions; one can also recall that $f(R)$ theories of gravity,
   widely applied to different problems of gravity and cosmology, can also be reduced to 
   STT with nonzero potentials $U(\phi)$. The papers \cite{clayton98,stab12,sta23,sta24a,sta24b}
   have made notable steps in this area, but much work is yet to be done. 
   The conformal mapping (2) makes it possible to use then the same methodology as in the
   present paper, considering the \pb\ equations in \ME\ and formulating the boundary 
   conditions in \MJ, but the study is complicated by the fact that analytical solutions in \ME\
   are known  only for some particular forms of the potential $U(\phi)$. For example, to our 
   knowledge, such solutions are unknown for a massive scalar field ($U(\phi) = \half m^2\phi^2$) 
   and even for $U(\phi) = \const$, which case is equivalent to a massless field in the presence 
   of a cosmological constant. Let us also recall that given such potentials in \ME, they look quite 
   differently in \MJ\ due to the conformal factor, see \eq (3).
  
   On the other hand, this study was restricted to canonical (non-phantom) scalar fields, whereas 
   many models containing phantom scalars are also of significant interest. In almost all cases
   (except those involving conformal continuations), the perturbation equations can be reduced 
   to those in the Einstein frame, but, as in the present paper, the appropriate boundary 
   conditions should be formulated ``individually'' for each solution. Though, solutions with 
   phantom fields are quite often singularity-free (wormholes, regular black holes etc.), and 
   the corresponding stability problems look quite differently from those described in this paper, 
   see, e.g., \cite{br-book,sarb1,stab11,stab12,stepan1,bk_uni,kb18} for reviews and discussions.

\Acknow 
     {The research of K. Bronnikov, S. Bolokhov and M. Skvortsova was supported by 
	RUDN University Project FSSF-2023-0003.
     F. Shaymanova and  R. Ibadov  gratefully acknowledge the support from
     Ministry of Innovative Development of the Republic of Uzbekistan, Project  No. FZ-20200929385.
}   
        


\small
\begin{thebibliography}{99}  

\bibitem{kon-zh11}
	R.A. Konoplya and A. Zhidenko,
	Quasinormal modes of black holes: from astrophysics to string theory,
	Rev. Mod. Phys. {\bf 83}, 793--836 (2011).
	
\bibitem{trap17}             
      K. A. Bronnikov. Trapped ghosts as sources for wormholes and regular black holes. 
      The stability problem. In: {\it Wormholes, Warp Drives and Energy Conditions}
      (Ed. F.S.N. Lobo, Springer, 2017), p. 137-160. 
	
\bibitem{stab18}             
	K.A. Bronnikov. Scalar fields as sources for wormholes and regular black holes,
	Particles {\bf 2018}, 1, 5; arXiv: 1802.00098.	

\bibitem{br-book}
	K. A. Bronnikov and S. G. Rubin.
     {\it Black Holes, Cosmology, and Extra Dimensions} (2nd edition, World Scientific, 2021).	
	
\bibitem{brito23}  
	Marco Brito, Carlos Herdeiro, Eugen Radu, Nicolas Sanchis-Gual, Miguel Zilhão,
	Stability and physical properties of spherical excited scalar boson stars,
     Phys. Rev. D 107, 084022 (2023); arXiv: 2302.08900. 

\bibitem{kb-hod}	
	K.A. Bronnikov and A.V. Khodunov. Scalar field and gravitational instability,
	Gen. Rel. Grav. {\bf 11}, 13 (1979).
	
\bibitem{fisher} 
	I. Z. Fisher, Scalar mesostatic field with regard for gravitational effects, 
	J. Eksp. Teor. Fiz. {\bf 18}, 636 (1948); gr-qc/9911008 (translation into English).     	
	
\bibitem{JNW} 
	A. I. Janis, E. T. Newman, and J. Winicour, Reality of the Schwarzschild singularity, 
	Phys. Rev. Lett. {\bf 20}, 878 (1968).

\bibitem{wyman81}
	Max Wyman, Static spherically symmetric scalar fields in general relativity,
	Phys. Rev. D {\bf 24}, 839–841 (1981).

\bibitem{penney}
     R. Penney,
     Generalization of the Reissner-Nordstr\"om solution to the Einstein field equations,
     Phys. Rev. {\bf 182}, 1383--1384 (1969).

\bibitem{STT1}
	P.G. Bergmann, Comments on the scalar-tensor theory,
	Int. J. Theor. Phys. {\bf 1}, 25 (1968).
	
\bibitem{STT2}
	R. Wagoner,
	Scalar-tensor theory and gravitational waves, \PRD {1} 3209 (1970).	
	
\bibitem{STT3}
	K. Nordtvedt, 
	Post-Newtonian metric for a general class of scalar-tensor gravitational 
	theories and observational consequences, \ApJ {161} 1059 (1970).

\bibitem{we23}
	K.A. Bronnikov, S.V. Bolokhov, M.V. Skvortsova, K. Badalov, R. Ibadov,
	On the stability of spherically symmetric space-times in scalar-tensor gravity,
	Grav. Cosmol. {\bf 29} (4), 374-386 (2023); arXiv: 2309.01794.

\bibitem{sadhu13}
	Amruta Sadhu and Vardarajan Suneeta,
	A naked singularity stable under scalar field perturbations,
     \IJMPD {22} 1350015 (2013); arXiv: 1208.5838.

\bibitem{chow20}
	Avijit Chowdhury and Narayan Banerjee, Echoes from a singularity,
     \PRD {102} 124051 (2020); arXiv: 2006.16522.	

\bibitem{clayton98}
	M. A. Clayton, L. Demopoulos, and J. L\'egar\'e, The dynamical
	stability of the static real scalar field solutions to the Einstein-Klein-Gordon 
	equations revisited, Phys. Lett. A {\bf 248}, 131 (1998).

\bibitem{sta23}
	O.S. Stashko, O.V. Savchuk, and V.I. Zhdanov,
	Quasi-normal modes of naked singularities in presence of non-linear scalar fields,
	Phys. Rev. D {\bf 109}, 024012 (2023); arXiv: 2307.04295.

\bibitem{sta24a}
	O. S. Stashko, O. V. Savchuk, and V. I. Zhdanov, Quasinormal modes of naked 
	singularities in presence of nonlinear scalar fields, 
	Phys. Rev. D {\bf 109}, 024012 (2024),

\bibitem{sta24b}
	V.I. Zhdanov, O.S. Stashko, and Yu.V. Shtanov,
	Spherically symmetric configurations in the quadratic $f(R)$ gravity,
	Phys. Rev. D {\bf 110}, 024056 (2024); arXiv: 2403.16741.

\bibitem{br73}
	K.A. Bronnikov, Scalar-tensor theory and scalar charge,
	Acta Phys. Pol. B {\bf 4}, 251 (1973).	
	
\bibitem{kb-CC2}
	K. A. Bronnikov, Scalar-tensor gravity and conformal continuations,
	 \JMP {43} 6096 (2002); gr-qc/0204001.
	 
\bibitem{br-star07}
     	K.A. Bronnikov and A.A. Starobinsky,
   	No realistic wormholes from ghost-free scalar-tensor phantom dark energy,
     	Pis'ma v ZhETF {\bf 85}, 1, 3-8 (2007);
     	JETP Lett. {\bf 85}, 1, 1-5 (2007). 
     	
\bibitem{skvo10}
    	K.A. Bronnikov, M.V. Skvortsova and A.A. Starobinsky,
    	Notes on wormhole existence in scalar-tensor and F(R) gravity.
    	\GC {16} 216 (2010); arXiv: 1005.3262.
	 	
\bibitem{ber-lei}
      O. Bergmann and R. Leipnik,
      Space-time structure of a static spherically symmetric scalar field,
      Phys. Rev. {\bf 107}, 1157 (1957).
      
\bibitem{h-ell73}
	H. Ellis, Ether flow through a drainhole: a particle model in general relativity,
	\JMP {14} 104 (1973).

\bibitem{BD-STT}
	C. Brans and R.H. Dicke, 
	Mach's principle and a relativistic theory of gravitation, Phys. Rev. {\bf 124} 925 (1961).

\bibitem{barker}
	B.M. Barker, General scalar-tensor theory of gravity with constant G,
	Astrophys. J. {\bf 219}, 5 (1978).
	
\bibitem{schwg}	
	J. Schwinger, {\it Particles, Sources and Fields} (Addison-Wesley, Reading, MA, Vol. 1, 1970).	
	
\bibitem{bruk94}
	William Bruckman, 	
	Generation of electro and magneto static solutions of the scalar-tensor theories of gravity,
	arXiv: gr-qc/9407003. 
	
\bibitem{sarb1}  	
	J.A. Gonz\'alez, F.S. Guzm\'an and O. Sarbach, 
	Instability of wormholes supported by a ghost scalar field. I. Linear stability analysis,
	Class. Quantum Grav. {\bf 26}, 015010 (2009); arXiv: 0806.0608.
 	
\bibitem{stab11}
     	K.A. Bronnikov, J.C. Fabris, and A. Zhidenko,
     	On the stability of scalar-vacuum space-times,
	Eur. Phys. J. C {\bf 71}, 1791 (2011).
	
\bibitem{stab12}
	 K.A.~Bronnikov, R.A.~Konoplya and A.~Zhidenko,
	 Instabilities of wormholes and regular black holes supported by a phantom scalar field,
	 Phys. Rev. D {\bf 86}, 024028 (2012); arXiv: 1205.2224.	
	
\bibitem{cold-q}
    K.A. Bronnikov, C.P. Constantinidis, R.L. Evangelista and J.C. Fabris,
    Electrically charged cold black holes in scalar-tensor theories,
    {Int. J. Mod. Phys.\/} {\bf D 8}, 481--505 (1999).

\bibitem{horn}
    T. Banks and M. O'Loughlin, 
    Classical and quantum production of cornucopions at energies below $10^{18}$ GeV,
    Phys. Rev. D {\bf 47}, 540 (1993).
    
\bibitem{kb-CC1}
     	K.A. Bronnikov, Scalar vacuum structure in general relativity and alternative theories.
	Conformal continuations,
	Acta Phys. Pol. B {\bf 32}, 3571 (2001); gr-qc/0110125.

\bibitem{BD-q}
	Maya Watanabe, A. W. C. Lun,
	Electrostatic potential of a point charge in a Brans-Dicke Reissner-Nordstrom field.
	Phys. Rev. D {\bf 88}, 045007 (2013); arXiv: 1305.6374.

\bibitem{bbm70}
	N.M. Bocharova, K.A. Bronnikov and V.N. Melnikov,
	On an exact solution of the Einstein-scalar field equations.
	Vestnik MGU Fiz., Astron. No. 6, 706 (1970).
	
\bibitem{turok92}
	B. Liu, L. McLerran and N. Turok, 
	Bubble nucleation and growth at a baryon-number-producing electroweak 
	phase transition, \PRD {46} 2668 (1992).
	 
\bibitem{bar-vis2}
	C. Barcel\'o and M. Visser, 
	Scalar fields, energy conditions, and traversable wormholes,
	\CQG {17} 3843 (2000); gr-qc/0003025.
	
\bibitem{stepan1}
	K.A. Bronnikov and S.V. Grinyok, 
	Conformal continuations and wormhole instability in scalar-tensor gravity,
	\GC {10} 237 (2004); gr-qc/0411063.	 
	
\bibitem{hybrid}	
	S. Capozziello, T.Harko, T. S. Koivisto, F. S. N. Lobo, and G. J. Olmo, 
	Hybrid metric-Palatini gravity,
	Universe {\bf 1}, 199 (2015); arXiv: 1508.04641.

\bibitem{g-hybrid}
	C. G. B\"ohmer and N. Tamanini, 
	Generalized hybrid metric-Palatini gravity, Phys. Rev. D {\bf 87}, 084031 (2013); arXiv: 1302.2355.

\bibitem{we-hyb}
	K. A. Bronnikov, S. V. Bolokhov, and M. V. Skvortsova, 
	Hybrid metric-Palatini gravity: Black holes, wormholes, singularities, and instabilities, 
	Grav. Cosmol. {\bf 26} (3), 212-227 (2020).

\bibitem{we-g-hyb}
	K. A. Bronnikov, S. V. Bolokhov, and M. V. Skvortsova, 
	Spherically symmetric space-times in generalized hybrid metric-Palatini gravity, 
	Grav. Cosmol. {\bf 27} (4), 358-374 (2021).
	
\bibitem{stab13}  
     K.A. Bronnikov, L.N. Lipatova, I.D. Novikov, and A.A. Shatskiy,
     Example of a stable wormhole in general relativity.
     \GC {19} 269--274 (2013); arXiv: 1312.6929.
	
\bibitem{od17-rev}
	S. Nojiri, S. D. Odintsov, and V. K. Oikonomou,
	Modified gravity theories in a nutshell: Inflation, bounce and late-time evolution,
	Phys. Rep. {\bf 692}, 1-104 (2017); arXiv: 1705.11098.

\bibitem{od08}
	S. Nojiri and S. D. Odintsov, 
	Modified non-local-F(R) gravity as the key for the inflation and dark energy,
	Phys. Lett. B {\bf 659}, 821 (2008).

\bibitem{kb-eliz10}
	K. A. Bronnikov and E. Elizalde, 
	Spherical systems in models of nonlocally corrected gravity,
	\PRD {81} 044032 (2010).

\bibitem{chrv22}
	S. V. Chervon, I. V. Fomin,  and A. A. Chaadaev,
	Spherically symmetric solutions of a chiral self-gravitating model in $f(R, \Box R)$ gravity,
	Grav. Cosmol. {\bf 28} (3), 295-303 (2022).

\bibitem{bk_uni}      
     K.A. Bronnikov, V.N. Melnikov and H. Dehnen,
     Regular black holes and black universes,
     \GRG {39} 973--987 (2007); gr-qc/0611022.

\bibitem{kb18}                   
	K.A. Bronnikov, Scalar fields as sources for wormholes and regular black holes.
	Particles {\bf 2018}, 1, 5; arXiv: 1802.00098.

\end{thebibliography}
\end{document}